\documentclass[a4paper, 12pt]{article}  
\usepackage{amsmath,amsthm,mathtools}
\usepackage{amsfonts}
\usepackage{amssymb}
\usepackage{comment}
\usepackage{a4wide}
\usepackage{jheppub}
\newtheorem{theo}{Theorem}
\newtheorem{coro}{Corollary}
\newtheorem{exa}{Example}
\newtheorem{lem}{Lemma}
\newtheorem{defi}{Definition}
\newtheorem{propo}{Proposition}
\newtheorem{remark}{Remark}
\DeclareMathOperator{\rk}{rank}
\DeclareMathOperator{\id}{\sc id}
\newcommand{\lra}{\longrightarrow}
\newcommand{\IQ}{\mathbb{Q}}
\newcommand{\IC}{\mathbb{C}}
\newcommand{\IZ}{\mathbb{Z}}
\newcommand{\IP}{\mathbb{P}}

\begin{document}

\preprint{Imperial-TP-2017-CH-03}

\title{Non-geometric Calabi-Yau Backgrounds and K3 automorphisms}
\author[a]{C.M. Hull,}
\author[b,c]{D. Isra\"el}
\author[d]{and A. Sarti}

\affiliation[a]{The Blackett Laboratory, Imperial College London, Prince Consort Road, London SW7 2AZ, United Kingdom}
\affiliation[b]{LPTHE, UMR 7589, Sorbonne Universit\'es, UPMC Univ. Paris 06, 4 place Jussieu, Paris, France}
\affiliation[c]{CNRS, UMR 7589, LPTHE, F-75005, Paris, France}
\affiliation[d]{Laboratoire de Math\'ematiques et Applications, UMR CNRS 7348, Universit\'e de Poitiers,
T\'el\'eport 2, Boulevard Marie et Pierre Curie, 86962 FUTUROSCOPE CHASSENEUIL, France}
\emailAdd{c.hull@imperial.ac.uk}
\emailAdd{israel@lpthe.jussieu.fr}
\emailAdd{Alessandra.Sarti@math.univ-poitiers.fr}

\null\vskip10pt
\abstract{We consider compactifications of type IIA superstring theory on mirror-folds obtained as K3 fibrations 
over two-tori with non-geometric monodromies involving mirror symmetries. At special points in the moduli space these are asymmetric Gepner models.
The compactifications are constructed  from non-geometric automorphisms that arise from the diagonal action of an 
automorphism of the K3 surface and of an automorphism of the mirror surface. 
We identify the corresponding gaugings of $\mathcal{N}=4$  supergravity in four dimensions, 
and show that the minima of the potential describe the same four-dimensional low-energy physics 
as the worldsheet formulation  in terms of asymmetric Gepner models. In this way, we obtain a class of Minkowski vacua of type II string theory which preserve $\mathcal{N}=2$  supersymmetry. The massless sector consists of $\mathcal{N}=2$ supergravity coupled
to 3 vector multiplets, giving  the STU model. 
 In some cases there are  additional massless hypermultiplets. 
}

                              
\maketitle

\section{Introduction}
Geometric compactifications  constitute only  a subset of string backgrounds and have interesting generalisations 
to non-geometric backgrounds. Examples arise from spaces with local fibrations that have
transition functions that include stringy duality symmetries. Spaces with torus fibrations and 
T-duality or U-duality transition functions are T-folds or U-folds~\cite{Hull:2004in}, 
while those with Calabi-Yau fibrations and mirror symmetry transition functions are mirror-folds~\cite{Hull:2004in}.
Such non-geometric spaces often have fewer moduli than their geometric counterparts, and 
the non-geometry typically breaks some of the symmetries, including supersymmetries, and provide 
an interesting tool for probing  quantum geometry. Solvable worldsheet conformal field theories (CFT's) such as 
asymmetric orbifolds can arise at special points in the moduli space of a non-geometric background~\cite{Dabholkar:2002sy}, allowing a 
complete analysis and important checks on general arguments.

Our focus here will be on mirror-folds of the type IIA superstring constructed 
from K3 bundles over $T^2$ with transition functions involving the mirror involution of the K3 surface. 
Previously Kawai et Sugawara have considered in~\cite{Kawai:2007nb} K3 mirror-folds with monodromies that, at 
least when the fiber is compact, break all supersymmetry; in the present work, we consider in contrast 
monodromies that preserve 8 supersymmetries,  i.e. which preserve a quarter of the 32 supersymmetries of 
the type IIA string, or a half of the 16 supersymmetries of type IIA compactified on $K3$. 
As a result, we find  interesting mirror-folds which give $D=4$, 
$\mathcal{N}=2$ Minkowski vacua of  type IIA superstring theory. As we shall see, particular examples 
give precisely the STU model of~\cite{Duff:1995sm} at low energies.

Our constructions can be viewed as particular cases of reductions with a duality twist~\cite{Dabholkar:2002sy}.
In such a construction, a theory in $D$ dimensions with discrete duality symmetry $G(\mathbb{Z)}$ (e.g. T-duality or U-duality) 
is compactified on a $d$-torus with a $G(\mathbb{Z)}$ monodromy around each of the $d$ circles, giving a string-theory 
generalization of Scherk-Schwarz reduction~\cite{Scherk:1979zr}. In many cases, the theory in $D$ dimensions 
has an action of the continuous group $G$ which is a symmetry of the low energy physics, 
but  which is broken to a discrete subgroup in the full string theory.
For a field $\phi$  transforming under $G$ as $\phi \mapsto g \phi$ the ansatz is
of the form
\begin{equation} 
\phi (x^\mu , y^i) = g(y) \hat \phi (x)
\end{equation}
where $y^i$, $i =1, \dots , d$,  are coordinates on $T^d$ and $x^\mu$, $\mu =0, \dots , D-d-1$,  are the remaining coordinates.
With periodicities $y^i\sim y^i + 2 \pi R_i$,
the mondromies
$g(y_i)^{-1} g(y_i + 2 \pi R_i)$ must be in $G(\mathbb{Z)}$ for each $i$.

Of particular interest are the special cases in which there are points in the moduli space in $D$ dimensions that 
happen to be fixed under the  action of the monodromies. 
For example, consider a theory where the moduli space in $D$ dimensions contains the moduli space for a 2-torus, $i.e.$ 
$SL(2,\mathbb{R})/U(1)$, identified under the action of the discrete group $SL(2,\mathbb{Z})$.
There are special points in the moduli space which are invariant under finite subgroups  of 
$SL(2,\mathbb{Z})$ isomorphic to $\mathbb{Z}_r$ for $r=2,3,4,6$. 
If one considers reductions on a circle where the monodromy is   in one 
of these  $\mathbb{Z}_r$ subgroups, then at 
any point in the moduli space invariant under the action of the monodromy,
 the reduction with a duality 
twist can be viewed as a  $\mathbb{Z}_r$ orbifold by a $\mathbb{Z}_r$ twist together with a shift around the 
circle by $2\pi R/r$~\cite{Dabholkar:2002sy}. From the effective field theory point of view, the moduli give scalar fields 
in $D-1$ dimensions and the reduction gives a potential for these fields. At each  fixed point in moduli space 
(i.e. at each point that is preserved by the monodromy)
the potential has a minimum at which it vanishes, giving a Minkowski compactification~\cite{Dabholkar:2002sy}.
In these cases, at the special points in moduli space there is both a stringy orbifold construction and 
a supergravity construction, giving complementary pictures of the same reduction.
This extends to a reduction on $T^d$ to $D-d $ dimensions with a duality twist on each of the $d$ circles.
The question of finding reductions giving Minkowski space in $D-d $ dimensions becomes related to that of 
finding fixed points in moduli space preserved by a subgroup of $G(\mathbb{Z)}$.

Our starting point will be the theory in six dimensions obtained from compactifying IIA 
string theory on a $K3$ surface. The moduli space of metrics and $B$-fields on $K3$ gives the moduli space of $K3$ CFT's
and is given by~\cite{Seiberg:1988pf,Aspinwall:1994rg}
\begin{equation*} 
\frac {O(4,20)}{O(4) \times O(20)} \, ,
\end{equation*}
identified under the discrete subgroup 
$O(\Gamma_{4,20})\subset O(4,20)$ preserving the lattice $\Gamma_{4,20}$, which is the lattice of total cohomology of the K3 surface 
as well as the lattice of D-brane charges in type IIA. $O(\Gamma_{4,20})$ 
is the perturbative duality group of two-dimensional conformal field theories 
with $K3$  target spaces, which includes mirror symmetries. 
We then reduce to four dimensions on $T^2$ with an $O(\Gamma_{4,20})$ twist around each circle.
Generically, such reductions  break all supersymmetry; we will  focus here on a class of monodromies admitting 
$\mathcal{N}=2$ vacua in four dimensions. From the string theory point of view, 
we will find them by considering points in the moduli space of sigma-models with K3 target spaces 
that are preserved by finite subgroups of $O(\Gamma_{4,20})$,  and taking orbifolds by such automorphisms combined with shifts on the circles.

The particular automorphisms of K3 CFTs that we will use are inspired by a worldsheet construction of asymmetric 
Gepner models presented in~\cite{Israel:2013wwa,Israel:2015efa}, following 
earlier works~\cite{Schellekens:1989wx,Intriligator:1990ua} 
(see~\cite{Blumenhagen:2016axv,Blumenhagen:2016rof} for later generalizations). They 
preserve only space-time supercharges from the worldsheet left-movers, and the asymmetry means that 
they are  non-geometric in general.

From the supergravity point of view, the conventional reduction on $T^2$  without twists gives $\mathcal{N}=4$ 
supergravity coupled to twenty-two $\mathcal{N}=4 $ abelian vector multiplets. The twisted reduction gives a gauged version of 
this supergravity with a non-abelian gauge group and a scalar potential. Vacua arise from minima of 
the potential, and of particular interest are theories 
with non-negative potentials and  minima that give zero vacuum energy 
 and a Minkowski vacuum.
Our construction gives string theory compactifications whose supergravity limits are of this type, and
moreover preserve $\mathcal{N}=2$ supersymmetry.

The relation between the asymmetric Gepner models  that underlie these compactifications and 
gauged supergravities was suggested by one of the authors in~\cite{Israel:2013wwa} 
and explored by Blumenhagen et al. in~\cite{Blumenhagen:2016axv} for 
a related but distinct class of models. In that work they considered 
quotients of   Calabi-Yau compactifications by   non-geometric automorphisms, 
rather than the freely-acting quotients involving torus shifts giving rise to fibrations over tori  that we consider here. 
In the freely-acting quotient, the scale of (spontaneous)  space-time supersymmetry breaking can be made arbitrarily small 
rather than being tied to the string scale as it would  be for quotients with fixed points. This means that for our constructions, 
the gauged supergravity approach gives a good description of 
the low-energy physics.
We will identify the gauging directly  from geometric considerations (rather than from identifying the massless spectra of the 
four-dimensional theories) and analyse the  worldsheet constructions of $K3$ 
mirror-folds with $\mathcal{N}=2$ supersymmetry from an algebraic geometry viewpoint. This approach 
will provide a powerful mathematical framework --~that applies to Calabi-Yau three-folds as well~-- and give an explicit construction of
the  low-energy four-dimensional gauged supergravity. 

We consider algebraic K3 surfaces defined as (the minimal resolution of) the zero-loci of quasi-homogeneous polynomials in 
weighted projective spaces $\mathbb{P}_{w_1,\ldots,w_n}$. These surfaces are characterized in particular by 
their {\it Picard number} $\rho$, the rank of their Picard lattice 
$S\, (X) = H^2 (X,\mathbb Z) \cap H^{(1,1)} (X)$ where $1\leqslant \rho \leqslant 20$. 
For the algebraic K3 surfaces of fixed $\rho$, the moduli space 
of CFTs factorizes into 
\begin{equation} 
\label{eq:mod_split}
\frac {O(2,20-\rho )}{O(2) \times O(20-\rho )}
\times \frac {O(2, \rho )}{O(2) \times O(\rho)}
\end{equation}
identified under a discrete subgroup, as we will  review in section~\ref{sec:math}. 
The first factor is interpreted as the complex structure moduli space of the algebraic surface 
and the second as the complex K\"ahler moduli space.

The definition of mirror symmetry for K3 surfaces is more subtle than for Calabi-Yau three-folds, 
as $K3$ is a hyperk\"ahler manifold. For algebraic K3 surfaces, the notion of {\it lattice-polarized mirror symmetry} 
(LP-mirror symmetry) was introduced around 40 years ago by Pinkham \cite{pinkham} and 
independently by Dolgachev and Nikulin \cite{ND1, ND2, ND3}; see the article by Dolgachev~\cite{Dolgachev:1996xw} 
for an introduction. In the special case considered by 
Aspinwall and Morrison~\cite{Aspinwall:1994rg},  this amounts to an $O(\Gamma_{4,20})$ transformation that maps  a K3 surface of Picard 
number $\rho$ to one with Picard number $20-\rho$, interchanging the two factors in eq.~(\ref{eq:mod_split}). Another notion 
of mirror symmetry, perhaps more familiar to physicists, is the Berglund-H\"ubsch construction~\cite{Berglund:1991pp} (BH mirror 
symmetry), generalizing the Greene-Plesser construction of mirror Gepner models~\cite{Greene:1990ud} 
to generic Landau-Ginzburg models with non-degenerate invertible polynomials. These two constructions (BH and LP mirror symmetry)
overlap but do not always agree.

A key ingredient to reconcile these two approaches to $K3$ mirror symmetry, which will also play a central role in the present study of   
non-geometric automorphisms, is to consider {\it non-symplectic automorphisms}~\cite{zbMATH03710310}, 
which are automorphisms  of the surface acting  on the 
holomorphic two-form $\omega$ as $\sigma_p^{\, \star} \, : \ \omega \mapsto \zeta_p\, \omega$, where $e.g.$ 
$\zeta_p = \exp 2i \pi /p$. 
As we will review in section~\ref{sec:math}, at least when $p$ is a prime number, 
the lattice-polarized mirror symmetry w.r.t. the invariant sublattice associated with the action of $\sigma_p$, coincides 
with the Berglund-H\"ubsch mirror symmetry~\cite{Artebani2014758,comparin2014}. An important corollary that we will obtain is that the automorphism $\sigma_p$ 
of the surface on the one-hand and the corresponding automorphism $\sigma_p^T$ of the (Berglund-H\"ubsch) 
mirror surface on the other hand act on sub-lattices of $\Gamma_{4,20}$ that are orthogonal to each other, 
denoted respectively  $T(\sigma_p)$ and $T(\sigma_p^T)$. It is expected that a similar statement is true 
for automorphisms of non prime order, 
and we expect the physics to work out in a similar fashion for such cases as well.

The non-geometric automorphisms of $K3$ CFTs that we study in this work correspond each to an 
$O(\Gamma_{4,20})$ transformation induced by a block-diagonal isometry in $O(T(\sigma_p) \oplus T(\sigma_p))$, the 
first block giving the isometry associated with the action of the automorphism $\sigma_p$ on the K3 surface, and the second block 
giving the isometry associated with the action of the automorphism $\sigma_p^T$ on the mirror K3 surface. These isometries of the lattice $\Gamma_{4,20}$ 
can be thought of as {\it mirrored automorphisms} of $K3$ CFTs of order $p$. They can be  decomposed as follows:
\begin{equation}
\hat{\sigma}_p := \mu ^{-1}  \circ \sigma_p^T \circ \mu  \circ \sigma_p \, ,
\end{equation}
where $\mu$ denotes the Berglund-H\"ubsch/lattice mirror involution. 
Here $\sigma_p $ is an order $p$ large diffeomorphism of $K3$, $\mu$  maps the $K3$ to its mirror, $\sigma_p^T $
is an order $p$ large diffeomorphism of  the mirror $K3$, and $ \mu ^{-1} $ maps the mirror $K3$ back to the original one.

Taking the quotient by two such automorphisms combined with shifts on the two one-cycles of a two-torus, 
at special points in the K3 moduli space fixed under the two automorphisms,  gives the 
asymmetric Gepner models of~\cite{Israel:2013wwa}. We extend this construction  to all points 
in moduli space using a reduction with duality twists. It is in general a difficult problem to 
find $O(\Gamma_{4,20})$ transformations that have fixed points in the K3 
moduli space and so can lead to Minkowski vacua. We will show in section~\ref{sec:nongeom} that the fixed 
points of the monodromies that we consider correspond indeed 
precisely to the Gepner model construction of~\cite{Israel:2013wwa} on the worldsheet (which 
are Landau-Ginzburg points in the moduli space of $K3$ CFTs with enhanced discrete symmetry) and lead to four-dimensional 
theories with $\mathcal{N}=2$ Minkowski vacua.

The type IIA string theory compactified on K3 is non-perturbatively dual to the heterotic string compactified on 
$T^4$~\cite{Hull:1994ys}. The duality symmetry group $O(\Gamma_{4,20})$ acts on the heterotic side through isometries 
of the Narain lattice, containing T-duality transformations as well as diffeomorphisms and shifts 
of the B-field, and is often referred to as the heterotic T-duality group.
The reduction from 6-dimensions on $T^2$ with $O(\Gamma_{4,20})$ monodromies 
round each circle  provides twisted reductions of precisely the type introduced and studied in~\cite{Dabholkar:2002sy}. 
These reductions can be regarded in general as T-fold reductions of the heterotic string, 
with transition functions involving the  T-duality group $O(\Gamma_{4,20})$~\cite{Hull:2004in}. At the 
special points of the moduli space that are fixed-points of the twists, the construction reduces to 
a reduction of asymmetric orbifold type, with a quotient by elements of the T-duality group $O(\Gamma_{4,20})$ 
combined with shifts on $T^2$~\cite{Dabholkar:2002sy}. 
Then the type IIA  K3 mirrorfolds are dual to heterotic T-folds, 
and at special points in the moduli space these become type IIA Gepner-type models and heterotic models of asymmetric 
orbifold type. Finding automorphisms with interesting fixed points is in general a difficult problem; 
the novelty here is that algebraic geometry leads us to a very interesting class of automorphisms that, 
when used in either the type IIA  or the heterotic description, gives a rich class of models with $\mathcal{N}=2$ supersymmetry.

This work is organized as follows. The first half of the paper, up to subsection~\ref{subsec:autLG}, is more algebraic-geometry 
oriented while the rest of the article deals with the physical aspects. In detail, section~\ref{sec:math} provides the 
necessary mathematical background about $K3$ surfaces, mirror 
symmetry and $K3$ automorphisms and section~\ref{sec:nongeom} presents the mirrored
automorphisms of mirror pairs of $K3$ surfaces and their relation with asymmetric Gepner model constructions. 
In section~\ref{sec:gauge}, we will consider the Scherk-Schwarz compactification of the 6-dimensional 
supergravity corresponding to type IIA compactified on $K3$ to obtain a four-dimensional gauged $\mathcal{N}=4$ supergravity. 
We show that for suitable choices of the twists, two gravitini become massive and two remain massless, giving vacua preserving $\mathcal{N}=2$ 
supersymmetry. In the final section, we will translate the $O(\Gamma_{4,20})$ monodromies defined in section~\ref{sec:nongeom} into 
the gauged supergravity framework, and analyze the moduli space of the low-energy theory.

\section{Mathematical background}
\label{sec:math}

The moduli space of two-dimensional conformal field theories defined by quantizing non-linear sigma models on K3 surfaces is given by~\cite{Seiberg:1988pf,Aspinwall:1994rg}:
\begin{equation} 
\label{eq:modsigma}
\mathcal{M}_\Sigma \cong O(\Gamma_{4,20}) \backslash O(4,20)/ O(4) \times O(20) \, ,
\end{equation}
where $O(\Gamma_{4,20})$ is the isometry group of the even and unimodular ($i.e.$ self-dual) lattice $\Gamma_{4,20}$ 
of signature $(4,20)$:
\begin{equation}
\Gamma_{4,20} \cong \Gamma_{3,19} \oplus U \, ,
\end{equation}
and
 $\Gamma_{3,19}$ 
is the K3 lattice
\begin{equation}
\Gamma_{3,19} \cong E_8 \oplus E_8 \oplus U \oplus U \oplus U\, ,
\end{equation}
which is isometric to the second cohomology group $H^{2}(X,\mathbb{Z})$ of a K3 surface $X$ endowed with its 
cup product.  
Here $U$ is the even unimodular lattice of signature $(1,1)$ and $E_8$ is the
even unimodular lattice of signature $(0,8)$ associated with  the Dynkin diagram $E_8$.

Automorphisms of $K3$ CFT's correspond  to isometries of the $\Gamma_{4,20}$ lattice. 
The geometric automorphisms of the surface form a subgroup
$O(\Gamma_{3,19}) \ltimes  \mathbb{Z}_{3,19}  \subset O(\Gamma_{4,20})$, generated by large 
diffeomorphisms of the surface in $O(\Gamma_{3,19}) \subset O(\Gamma_{4,20})$
and transformations $ \mathbb{Z}_{3,19}$ corresponding to 
shifts of the B-field by representatives of integral cohomolgy classes.
 Isometries that are not in 
$O(\Gamma_{3,19}) \ltimes  \mathbb{Z}_{3,19}$ are \lq non-geometric'.

An important sublattice of the K3 lattice $\Gamma_{3,19}$ of a K3 surface $X$ is the {\it Picard lattice}\footnote{
The Picard group or Picard lattice is generated by algebraic curves  of the surface, 
$i.e.$ curves that are holomorphically embedded in $X$.} 
which is defined to be:
\begin{equation}
S\, (X) = H^2 (X,\mathbb Z) \cap H^{(1,1)} (X) \, .
\end{equation}
The rank of this lattice ($i.e.$ the rank of the corresponding Abelian group) $\rho (X)$, 
or Picard number, is at least one for any algebraic 
K3 surface, and its signature is  $(1,\rho-1)$. The  {\it transcendental lattice} $T(X)$
of an algebraic K3 surface $X$
is defined to be the  sub-lattice of $H^2 (X,\mathbb Z) \cong \Gamma_{3,19}$ 
orthogonal to the Picard lattice:
\begin{equation}
T (X) = H^2(X,\IZ) \cap S \, (X )^{\perp}\, \hookrightarrow  \Gamma_{3,19} \,.
\end{equation}
This lattice has signature $(2,20-\rho)$.

For sigma-model CFTs on algebraic K3 surfaces, 
the Picard lattice can be enlarged to the \lq quantum Picard lattice':
\begin{equation}
S^{\mathfrak{Q}}  (X ) \cong S\, (X) \oplus U\, .
\end{equation}
with signature $(2,\rho)$.
The transcendental lattice 
can also be viewed as
the orthogonal complement of the quantum Picard lattice, i.e.
$T (X) = \Gamma_{4,20} \cap S^{\mathfrak{Q}}  (X )^{\perp}$. The 
moduli space of non-linear sigma-model CFTs on algebraic K3 surfaces with a Picard lattice of Picard number $\rho$ then factorizes 
as~\cite{Aspinwall:1994rg}:
\begin{equation}
\label{eq:modrho}
\mathcal{M}_\Sigma^\rho \cong  O(T(X)) \backslash O(2,20-\rho)/ (O(2)\times O(20-\rho))
\ \times \ 
O(S^{\mathfrak{Q}}(X)) \backslash O(2,\rho) / (O(2)\times O(\rho)).
\end{equation}
One may think of the first factor as corresponding to the complex structure moduli space and 
the second one to the complex K\"ahler moduli space.

Mirror symmetry of Calabi-Yau three-folds exchanges their Hodge numbers 
$h^{2,1}$ and $h^{1,1}$, and exchanges the complex structure and complex 
K\"ahler moduli spaces. For $K3$ surfaces the situation is different since $(i)$ all $K3$ surfaces 
are diffeomorphic to each other and so have 
the same Hodge numbers and topology,  and $(ii)$ as  these manifolds are hyperk\"ahler, 
the complex structure and complex K\"ahler moduli are not unambiguously defined. 
For algebraic $K3$ surfaces there are two different notions of mirror symmetry 
that we will review in turn below, and these will both play a role 
in the construction of the non-geometric automorphisms we use in this paper.

\subsection{Lattice-polarized mirror symmetry}

A first notion of mirror symmetry of algebraic $K3$ surfaces is the {\it lattice-polarized} (LP) mirror symmetry discussed by 
Dolgachev in~\cite{Dolgachev:1996xw} and introduced by Pinkham~\cite{pinkham}, 
Nikulin and Dolgachev~\cite{ND1, ND2, ND3}. This construction uses the embedding of a given lattice $M$ as a sub-lattice of 
the Picard lattice, which should be primitive. A primitive embedding of a lattice $M$ into a lattice $N$, 
$\iota \, : \ M \hookrightarrow N$, is such that, viewing $N$ as an Abelian group and $\iota (M)$ as a subgroup, 
the quotient $N/\iota(M)$ is a torsion-free Abelian group.\footnote{As a counter-example, 
if  $(e_1 , e_2)$ is a basis of $N$, and if $\iota (M)$ is 
spanned by $\{ e_1 + e_2 , e_1 -e_2 \}$, the embedding $\iota \, : \  M \hookrightarrow N$ 
is not primitive as $N/\iota(M) \cong \mathbb{Z}_2$.}

\begin{defi}
Let $M$ be an even lattice of rank $t+1$ and signature $(1,t)$, with $t\leqslant 18$, admitting a 
primitive embedding in the K3 lattice, $\iota \, : \ M \hookrightarrow \Gamma_{3,19}$. Assuming that 
its orthogonal complement $\iota (M)^\perp \subset \Gamma_{3,19}$ admits a primitive embedding 
$\iota ' \, : \ U \hookrightarrow \iota (M)^\perp$,  the  mirror lattice $M^\vee$ to $M$ of rank $19-t$ and signature $(1,18-t)$
is defined through the decomposition 
\begin{equation}\label{mirror}
\iota (M)^\perp = \iota ' (U) \oplus M^\vee \, .
\end{equation}
If 
 there exists a primitive embedding  $\jmath\, : \ M \hookrightarrow S (X)$ of 
$M$ in the Picard lattice of an algebraic $K3$ surface $X$, then we say that $X$ is an $M$-polarized K3 surface. 
Then two K3 surfaces $X$ and $X^{\vee}$ form a  lattice  mirror pair
if $X$ is $M$-polarized and $X^{\vee}$ is $M^{\vee}$-polarized with $M^{\vee}$ the mirror lattice to $M$.
\end{defi}
  
In other words, the LP mirror construction  associates 
each $M$-polarized K3 surface with an $M^\vee$-polarized K3 
surface, with  $\text{rk}\, (M^\vee )= 20 - \text{rk}\, (M)$. The moduli space of 
the family of 
$M$-polarized K3 surfaces and the moduli space of 
the family of  $M^\vee$-polarized K3 
surfaces are called {\it LP mirror moduli spaces}.

\begin{exa}
\normalfont
The Aspinwall-Morrison construction of mirror symmetry~\cite{Aspinwall:1994rg} is a particular instance of 
LP mirror symmetry. The authors 
considered an algebraic $K3$ surface $X$  polarized by the whole Picard lattice, $i.e.$ with $M$ embedded 
primitively  as $\iota (M) = S(X)$. One 
has then 
\begin{equation}
S (X)^\perp = T(X) = \iota ' (U) \oplus M^\vee \, ,
\end{equation}
and $M^\vee$ is the Picard lattice of the mirror surface. In other words, mirror symmetry is defined as the map 
\begin{equation}
X \mapsto\ X^\vee \, ,
\end{equation}
where
\begin{equation}
T (X) =S^{\mathfrak{Q}} (X^\vee), \qquad S^{\mathfrak{Q}}(X)= T(X^\vee) \, ,
\end{equation}
which exchanges the quantum Picard lattice and the transcendental lattice. 
Hence both factors in~(\ref{eq:modrho}) are exchanged under this involution, which 
can be viewed as exchanging the complex structure and the complex K\"ahler moduli spaces of sigma-model CFTs 
on the surface. 
\end{exa}

\subsection{Berglund-H\"ubsch mirror symmetry}

The second notion of mirror symmetry, {\it Berglund-H\"ubsch (BH) mirror symmetry}, is not specific to K3 surfaces. It 
follows from the Greene-Plesser construction~\cite{Greene:1990ud} of mirror Gepner models~\cite{Gepner:1987qi} discovered 
in physics, exchanging the vector and axial $R$-currents of the worldsheet $(2,2)$ superconformal field theories. It was generalized  
by Berglund and H\"ubsch~\cite{Berglund:1991pp} and Krawitz~\cite{Krawitz2010}; later Chiodo and Ruan proved 
in~\cite{CHIODO20112157} that it coincides with  cohomological mirror symmetry.

Let us consider a K3 surface realized as the minimal resolution of a hypersurface in  a weighted projective space 
$\mathbb{P}_{[w_1 w_2 w_3 w_4]}$ with $\text{gcd}\, (w_1,\ldots,w_4)=1$. 
A polynomial $W\, : \ \mathbb{C}^4 \to \mathbb{C}$ is quasi-homogeneous of degree $d$ if 
\begin{equation}
\forall \lambda \in \mathbb{C}^* \ , 
\quad W(\lambda^{w_1} x_1,\ldots, \lambda^{w_4} x_4) = \lambda^d  W( x_1,\ldots, \ x_4)\, .
\end{equation}
It is non-degenerate if the origin $x_1=\cdots = x_4 =0$ 
is the only critical point and if the fractional 
weights $w_1/d,\ldots,w_4/d$ are uniquely determined by $W$. If furthermore the 
number of monomials equals the number of variables 
the polynomial is said to be {\it invertible}. By rescaling one can then put  an invertible polynomial $W$ in the form 
\begin{equation}
W = \sum_{i=1}^4 \prod_{j=1}^4 \, {x_{j}}^{\ a^i_{\, j}}\, ,
\end{equation}
where the square matrix $A_W := ( a^i_{\, j})$ is invertible.  If $\sum_{\ell=1}^4 w_\ell = d$, 
the  hypersurface $\{ W=0 \}$ in $\mathbb{P}_{[w_1 w_2 w_3 w_4]}$ admits 
a minimal resolution $X_W$ which is a smooth K3 surface.

We denote by $G_W$ the Abelian group of all diagonal scaling transformations preserving the polynomial $W$:
\begin{equation}
G_W = \{ (\mu_1,\ldots,\mu_4) \in (\mathbb{C}^\star)^4 \, | \, \ W(\mu_1 x_1, \ldots \mu_4 x_4) = W(x_1, \ldots  x_4) \}\, ,
\end{equation}
and $SL_W$ its subgroup containing  elements of the form 
\begin{equation}
(\mu_1 ,\ldots, \mu_4 ) = \left(e^{2i\pi g_1},\ldots,e^{2i\pi g_4}\right) \ ,\quad \sum_{\ell=1}^4 g_\ell \in \mathbb{Z} \, , 
\end{equation}
which corresponds in physics to the group of supersymmetry-preserving symmetries. 
The group $G_W$ always contains   the element $j_W$ with $(g_1=w_1/d,\ldots,g_4 = w_4 /d)$ generating 
a cyclic group $J_W$ of order $d$.

Let us consider a subgroup $G \subset G_W$ such that  $J_W \subseteq G \subseteq SL_W$,  and the quotient group 
$\tilde{G} = G/J_W$. The minimal resolution of the orbifold $X_W /\tilde{G}$, denoted $X_{W,G}$, 
is also a K3 surface, that we associate with the pair $(W,G)$. 
In physics, an orbifold of a $K3$ sigma-model (or Landau-Ginzburg model) by a discrete group $G$ satisfying 
the condition $J_W \subseteq G \subseteq SL_W$ preserves all space-time supersymmetry.  

We now introduce the Berglund-H\"ubsch mirror symmetry, which follows in the physical context from the 
isomorphism between the superconformal field theories associated 
with a pair of Landau-Ginzburg orbifolds, generalizing the original Greene-Plesser construction of mirror Gepner model orbifolds.

\begin{defi}
Let $(W,G)$ be associated with the minimal resolution of $X_W /\tilde{G}$, a smooth K3 surface. 
The pair $(W^T,G^T)$ is obtained as follows:
\begin{itemize}
\item $W^T$ is specified by the matrix $A_{W^T} : = (A_W)^T$. 
\item $G^T = \{ g \in G_{W^T} \, , \ g A_W h^T \in \mathbb{Z}\, , \ \forall h \in G \}$. 
\end{itemize}
The Berglund-H\"ubsch 
mirror surface of $X_{W,G}$  is then given by $X_{W^T,G^T}$, the minimal resolution of the orbifold $X_{W^T} /\widetilde{G^T}$.
\end{defi}
\noindent
Here $X_{W^T} $ is the surface $W^T(\tilde x^1, \dots \tilde  x^4)=0$
and  $g= (g_1,\dots, g_4)$ 
is an automorphism of $X_{W^T} $ acting on the 
coordinates $\tilde x^\ell$  as $\tilde x^\ell \mapsto \exp {(2\pi i \, g_\ell)} \, \tilde x^\ell$
for $\ell=1,2,3,4$ while
$h= (h_1,\dots, h_4)$ specifies an automorphism of $X_W$ 
under which the coordinates scale as $  x^\ell \mapsto \exp {(2\pi  i \, h_\ell)} \,  x^\ell$.
It is straightforward to check that, if the pair $(W,G)$ is associated with a 
smooth $K3$ surface obtained as the minimal resolution of $X_W /\tilde{G}$,  then
 the mirror pair $(W^T,G^T)$ is associated with a smooth $K3$ 
surface obtained as the minimal resolution of $X_{W^T} /\widetilde{G^T}$, as  
$J_{W^T} \subseteq G^T \subseteq SL_{W^T}$.

\begin{exa}
\normalfont
A Fermat-type K3 surface in a weighted projective space is defined by the polynomial 
\begin{equation}
W = x_1^{\, d/w_1}  + x_2^{\, d/w_2} + x_3^{\, d/w_3} + x_4^{\, d/w_4}\, ,
\end{equation}
where $d = \text{lcm} \, (w_1, \ldots, w_4)$.
This polynomial is preserved by the symmetries under which any of the coordinates scales as
$x^\ell \mapsto  \exp {(2\pi i  \, w_\ell /d)} x^\ell$ and the other coordinates are invariant. They generate the 
group of diagonal symmetries 
$G_W \cong \mathbb{Z}_{d/w_1}
\times
\mathbb{Z}_{d/w_2}
\times
\mathbb{Z}_{d/w_3}
\times
\mathbb{Z}_{d/w_4}$. 
Consider the case in which we choose the group $G$ to be $J_W = 
\langle  ( w_1/d, \ldots, w_4/d)\rangle$. 
Then $\tilde G$ is the trivial group  and $X_{W,G}=X_W$. 
The mirror surface is characterized by the same polynomial as $W^T=W$, and the dual group 
$G^T$
is given by the elements 
$$g = (g_1,\ldots,g_4) \in G_{W^T}=G_{W}
$$
satisfying the condition
\begin{equation} 
r \sum_{\ell=1}^4 g_\ell \in \mathbb{Z}\, , \forall r \in \{0,\ldots,d-1\}\, .
\end{equation}
As $g_\ell = n_\ell \frac{w_\ell}{d}$ for some integer $n_\ell$, we get the condition
\begin{equation}
\sum_{\ell=1}^4  \frac{w_\ell}{d} \, n_\ell \in \mathbb{Z} \, .
\end{equation}
This means that $G^T = SL_W$ in this case. 
\end{exa}

\subsection{Non-symplectic automorphisms of prime order}
\label{subsec:prime}

The two notions of mirror symmetry of algebraic $K3$ surfaces do not necessarily agree. 
In particular, as was shown in~\cite{doi:10.1081/AGB-120006479}, the 
LP mirror of a surface polarized by its whole Picard lattice is not always 
identical to the BH mirror of the same surface. 
There exists nevertheless a class of lattice-polarized mirror symmetries, 
in which the surface is polarized by a sub-lattice of the Picard lattice, that gives the same results 
as the Berglund-H\"ubsch construction, and that will be instrumental in our construction of 
non-geometric automorphisms.

Let us first define a non-symplectic automorphism of order $p$ of a $K3$ surface $X$ as a diffeomorphism $\sigma_p:X\to X$
of the surface acting  on the holomorphic two-form $\omega (X)$ as 
\begin{equation}
\sigma_p^{\, \star} \, : \ \omega(X) \mapsto \zeta_p\, \omega(X)\, , 
\end{equation} 
where $\zeta_p$ is a primitive $p$-th root of unity, $i.e.$ such that $\zeta_p$ generates 
a cyclic group isomorphic to $\mathbb{Z}/p\mathbb{Z}$, $e.g.$ $\zeta_p=\exp (2i\pi/p)$.\footnote{These autormorphisms 
are often called in the literature {\it purely non-symplectic} but for simplicity well call them just non-symplectic. 
When the order $p$ is a primer number, as it is the case in the present work, each 
non-symplectic automorphism is purely non-symplectic.} 
If $p$ is a prime number then it is straightforward to see that $2\leq p\leq 19$ (see~\cite[Theorem 0.1]{zbMATH03710310}).

The automorphism $\sigma_p:X\to X$ acts on 2-forms through $\sigma_p^{\, \star}$.
Let $S(\sigma_p)$ be the sub-lattice of $H^2 (X,\mathbb{Z})$ invariant under the action of 
the isometry $\sigma_p^{\, \star}$  and $T(\sigma_p)$ its orthogonal complement. 
The rank of the lattice $S(\sigma_p)$ will be denoted by $\rho_p$. 
As was shown by Nikulin~\cite{zbMATH03710310},  the invariant sublattice $S(\sigma_p) $ is a subset of the Picard lattice, 
\begin{equation}
S(\sigma_p) \subseteq S(X)\, 
\end{equation}
and both  $S(\sigma_p)$  and $T(\sigma_p)$ are primitive sub-lattices of the $K3$ lattice.
The following lemma was also proved in~\cite{zbMATH03710310}: 
\begin{lem}
\label{lem:diag}
Let $\sigma_p$ be an order $p$ non-symplectic automorphism of a K3 surface $X$. There 
exists a positive integer $q$ such that the action 
$\sigma^\star_p $ on the vector space $T (\sigma_p) \otimes \mathbb{C}$ can be diagonalized as 
\begin{equation}
\label{diago}
\begin{pmatrix}
\zeta_p \mathbb{I}_q & 0& \cdots & \cdots & \cdots &0\\
0 &  \ddots & & &  &\vdots\\
\vdots & & & \zeta_p^n \mathbb{I}_q & & \vdots\\
\vdots& & & &\ddots & 0 \\
0 & \cdots & \cdots & \cdots & 0 & \zeta_{p}^{p-1} \mathbb{I}_q
\end{pmatrix}
\end{equation}
where $\mathbb{I}_q$ is the identity matrix in $q$ dimensions and all  integers 
$n \in \{ 1,p-1\}$ with $\mathrm{gcd}\, (n,p)=1$ appear once, $i.e.$ all primitive $p$-roots of unity 
are eigenvalues and the corresponding eigenspaces are all of dimension $q$. 
\end{lem}
\begin{remark}\label{unicity} 
{\rm 
From~\cite[Proposition 9.3]{AST} and \cite{zbMATH03710310} the action of the non-symplectic automorphism on the K3 lattice
is unique up to conjugation with isometries.
} 
\end{remark}

We will consider a particular type  of hypersurface in weighted projective space
admitting a non-symplectic automorphism of prime order $p$, whose non-degenerate invertible polynomial is 
of the form
\begin{equation}
\label{eq:pcycl}
W = x_1^{\, p} + f(x_2,x_3,x_4)\, . 
\end{equation}
We will call such surfaces {\it p-cyclic}, following~\cite{comparin2014}. They admit the obvious order $p$ non-symplectic 
automorphism  $\sigma_p \, : \ x_1 \mapsto \zeta_p\,  x_1$. By construction, the BH mirror of a $p$-cyclic surface 
has its  defining polynomial of the form $W^T=\tilde{x}_1^{\, p} + \tilde{f}(\tilde{x}_2,\tilde{x}_3,\tilde{x}_4)$, 
therefore it also  admits an order $p$ automorphism $\sigma_p^{\, T} \, : \ \tilde{x}\mapsto \zeta_p \, \tilde{x}$. 

The following theorem was proved for $p=2$ by
Artebani et al.~\cite{Artebani2014758} and for 
$p\in \{3,5,7,13 \}$ by Comparin et al.~\cite{comparin2014}.\footnote{As shown by Nikulin~\cite{zbMATH03710310}, K3 surfaces 
admitting non-symplectic automorphisms of prime order up to $p=19$ exist; 
however for the values $p\in \{ 11,17,19 \}$ one cannot present the surface in a $p$-cyclic form, as was noticed in~\cite{comparin2014}.}

\begin{theo}
\label{theo:mir}
Let $X_{W,G}$ be a $p$-cyclic $S(\sigma_p)$-polarized K3 surface, 
where $S(\sigma_p)\subseteq S(X)$ is the sub-lattice of the Picard lattice invariant under the action of the 
non-symplectic automorphism $\sigma_p$ of prime order, with $p\in \{2,3,5,7,13 \}$. 
Let  $X_{W^T,G^T}$ be its Berglund-H\"ubsch mirror, polarized by the 
invariant sublattice $S(\sigma_p^{\, T})$ associated with the non-symplectic automorphism $\sigma_p^{\, T}$. 
Then  $X_{W,G}$ and $X_{W^T,G^T}$ belong to mirror families of $K3$ surfaces in the sense of lattice-polarized mirror symmetry.
\end{theo}
We obtain from this theorem a simple corollary which will play an important role in the 
construction of non-geometric compactifications. We first define the quantum invariant sublattice as the orthogonal 
complement of $T(\sigma_p)$ in the $\Gamma_{4,20}$ lattice, namely 
\begin{equation}
S^\mathfrak{Q} (\sigma_p) \cong S (\sigma_p) \oplus U\, ,
\end{equation}
which has signature $(2,\rho_p)$ with $\rho_p \leqslant \rho (X)$. For a lattice $L$ we denote by $L^{\mathbb{R}}$ the 
real vector space $L\otimes \mathbb{R}$  generated by its basis vectors. 

 \begin{coro}
\label{coro:decomp}
Let $X_{W,G}$ be a $p$-cyclic  $S(\sigma_p)$-polarized K3 surface, 
where $S(\sigma_p)$ is the invariant sublattice under the $\sigma_p$ action, with $p\in \{2,3,5,7,13 \}$.  
and $X_{W^T,G^T}$ its LP mirror,  regarded as an $S(\sigma_p^T)$-polarized K3 surface, 
which is also its Berglund-H\"ubsch mirror following 
Theorem~\ref{theo:mir}. 

From the theorem we have $T(\sigma_p)= S(\sigma_p)^\perp \cap \Gamma_{3,19} = U \oplus S(\sigma_p^T)$ and similarly  
$T(\sigma_p^T)= S(\sigma_p^T)^\perp\cap \Gamma_{3,19}  = U \oplus S(\sigma_p)$. Hence $T(\sigma_p^{\, T})$ 
is the orthogonal complement\footnote{The embedding of $T(\sigma_p)$ may not be unique in $\Gamma_{4,20}$ 
but we tacitly choose the embedding such that $T(\sigma_p^T)$ is the orthogonal complement.}
 of 
$T(\sigma_p)$ in $\Gamma_{4,20}$:
\begin{equation}
T(\sigma_p^{\, T}) \cong  T(\sigma_p)^\perp \cap \Gamma_{4,20}\, .
\end{equation}
We obtain then the orthogonal decomposition over $\mathbb{R}$:
\begin{equation}
\label{eq:latticedecomp}
\Gamma^{\mathbb{R}}_{4,20} \, \cong \, T(\sigma_p^T)^{\mathbb{R}} \oplus T(\sigma_p)^{\mathbb{R}}\, .
\end{equation}
\end{coro}

\begin{exa}
\label{exa:selfmirror}
\normalfont
In order to illustrate this general construction, we consider first the self-mirror K3 surface $X$ given by the hypersurface
\begin{equation}
\label{eq:self_mirr}
w^{2} + x^{3} + y^{7} + z^{42}=0
\end{equation}
in the weighted projective space $\mathbb{P}_{[21,14,6,1]}$. It inherits the orbifold 
singularities of the ambient space, and these should be minimally resolved in order to obtain a smooth K3 surface. 
This K3 surface 
admits a non--symplectic automorphism of order 42, acting on $z$ by $$
z \mapsto e^{2i\pi/42} z 
$$
and leaving  the other coordinates invariant.
Then this implies, by~\cite[Theorem 0.1]{zbMATH03710310}, that
the rank of the transcendental lattice is a multiple of the Euler function of $42$, which is $12$. Since the rank is necessarily less than $22$, 
which is the rank of the K3 lattice, the rank  is necessarily equal to $12$ and the rank of the Picard lattice is then $10$. 
Interestingly, by~\cite{doi:10.1081/AGB-120006479} the generic K3 surface in the weighted projective space  $\mathbb{P}_{[21,14,6,1]}$ has
Picard lattice of rank 10, which is the same as the rank of the Picard lattice of the surface~\eqref{eq:self_mirr} of Fermat type.
The Picard lattice $S(X)$  is isometric to a self dual lattice of signature $(1,9)$, which is
\begin{equation}
S(X) \cong E_8 \oplus U\, .
\end{equation}
Thus this surface is its own mirror in the sense of Aspinwall-Morrison, as we have
\begin{equation}
\label{eq:transc}
S^\mathfrak{Q}(X) \cong T (X) \cong E_8 \oplus U \oplus U\, .
\end{equation}

The moduli space of complex structures associated with this surface corresponds 
to the set of space-like two-planes in $\mathbb{R}^{3,19}$ orthogonal to the 
basis vectors of $S (X)$, quotiented by $O(T(X))$, the group of isometries of the transcendental lattice:
\begin{equation}
\mathcal{M}_\text{cs} \cong O(T(X))\backslash O(2,10)/O(2)\times O(10)
\end{equation}
of real dimension 20.

The hypersurface~(\ref{eq:self_mirr}) admits several non-symplectic automorphisms of prime order. 
The hypersurface is $p$-cyclic for $p=2,3,7$ and the corresponding automorphisms 
$\sigma_2,\sigma_3,\sigma_7$ are of order 2, 3 and 7. Their action is 
\begin{align}
\sigma_2\, &: w \mapsto - w\, , \notag\\
\sigma_3\, &: x \mapsto e^{2i\pi/3} x\, , \notag\\
\sigma_7\, &: y \mapsto e^{2i\pi/7} y\, .
\end{align}
In all cases,   the invariant 
sublattice $S(\sigma_p) \subset \Gamma_{3,19}$ is identified with the Picard lattice $S(X)$ (see~\cite{AST,comparin2014}), and 
$S^\mathfrak{Q} (\sigma_p) \cong S^\mathfrak{Q} (X)$. Its orthogonal 
complement in $\Gamma_{4,20}$ is naturally the transcendental lattice of the surface, hence $T(\sigma_p)\cong T(X)$.

The action of $\sigma_3$ on the transcendental lattice~(\ref{eq:transc}) of 
the surface~(\ref{eq:self_mirr}) is given, in the appropriate basis over $\IC$, by six copies of the 
companion matrix of the cyclotomic polynomial $\Phi_3=\prod_{n=1}^2(z-e^{2i\pi n/3})$, using Lemma~\ref{lem:diag}.\footnote{
The companion matrix of a polynomial of the form $P(x) = a_0+a_1 x +\cdots + a_{n-1} x^{n-1} + x^n$ is the matrix 
{\tiny $\left( \begin{array}{ccccc}0& 0 &\cdots&0& - a_0 \\ 1&0 & \cdots & 0 & -a_1 \\ 0&1 & \cdots & 0 & -a_2 \\ 
\vdots & \vdots & \ddots & \vdots & \vdots \\ 0&0 & \cdots & 1 & -a_{n-1} \end{array} \right)$}, 
whose characteristic polynomial is $P$.}

Recall that by Remark \ref{unicity} the action can be given in a unique way on the transcendental lattice. 
It splits into an action onto the $U\oplus U$ lattice and onto the $E_8$ lattice. For $U\oplus U$ we
choose a lattice basis in which the lattice metric (or Gram matrix) is 
\[\arraycolsep=2.6pt
\label{eq:gramuu}
\left( \begin{array}{cccc}
 0 & 1 & 0 & 0 \\
 1 & 0 & 0 & 0 \\
 0 & 0 & 0 & 1 \\
 0 & 0 & 1 & 0 \\
\end{array}\right)\, . 
\]
The action of the isometry $\sigma_3$ on  $U\oplus U$  is then given by 
the following matrix in $ O(U\oplus U)$:
\begin{equation}
\label{eq:j3uu}
M_3^{U\oplus U}=\begin{pmatrix}
 1 & 0 & 1 & 0 \\
 0 & -2 & 0 & 3 \\
 -3 & 0 & -2 & 0 \\
 0 & -1 & 0 & 1 \\
\end{pmatrix}\, .
\end{equation}
For the $E_8$ lattice, choosing a lattice basis in which the lattice metric is
\begin{equation}
\label{eq:gram8}
E_8^{st} :=
\begin{pmatrix}
 -2&1&0&0&0&0&0&0\\
1&-2&1&0&0&0&0&0\\
0&1&-2&1&0&0&0&0\\
0&0&1&-2&1&0&0&0\\
0&0&0&1&-2&1&1&0\\
0&0&0&0&1&-2&0&0\\
0&0&0&0&1&0&-2&1\\
0&0&0&0&0&0&1&-2 
\end{pmatrix} \, ,
\end{equation}
the action of $\sigma_3$ on  $E_8$  is given by the following matrix in $O(E_8)$:
\begin{equation}\label{eq:j3e8}
M_3^{E_8} :=
\begin{pmatrix}
0&0&0&0&0&-1&1&-1\\
1&-1&1&-1&0&-1&2&-2\\
2&-1&1&-1&0&-1&2&-3\\
2&-1&2&-1&-1&-1&3&-4\\
3&-2&2&0&-2&-1&4&-5\\
2&-1&1&0&-1&-1&2&-2\\
2&-2&2&0&-1&-1&2&-3\\
1&-1&1&0&0&-1&1&-2
\end{pmatrix}\, .
\end{equation}

Likewise, the action of $\sigma_7$ is given in the appropriate basis 
over $\IC$ by two copies of the companion matrix of the cyclotomic polynomial 
$\Phi_7=\prod_{n=1}^6(z-e^{2i\pi n/7})$, and finally the action of $\sigma_2$ 
on $T(X)$ is simply given by minus the identity matrix in twelve dimensions.

For this surface $|SL_W/J_W|=1$ hence $J_{W^T}=J_W$. Therefore, using either of the 
non-symplectic automorphisms of prime order, one finds 
that the surface $X_{W,J_W}$ is its own Berglund-H\"ubsch mirror and, polarized by $S(\sigma_p) \cong S(X)$, is also its 
own mirror in the sense of LP mirror symmetry.

\end{exa}

\begin{exa}\label{exa:autre}
\normalfont
Let us consider the hypersurface
\begin{equation}
\label{eq:23824}
w^{\, 2} + x^{\, 3} + y^{\, 8} + z^{\, 24}=0
\end{equation}
in the weighted projective space $\mathbb{P}_{[12,8,3,1]}$. In this case $|SL_W/J_W|=2$ and 
there are two choices of $G$ with $J_W \subseteq G \subseteq  SL_W$, either $G=J_W$ or $G=SL_W$. Berglund-H\"ubsch mirror 
symmetry provides then the mirror pair $(W,J_W)$ and $(W,SL_W)$.

The surface defined by eq.~(\ref{eq:23824}) admits  a non-symplectic automorphism of order 3 acting 
as $\sigma_3 \, : \ x \mapsto e^{2i\pi/3} x$, 
while keeping the other variables fixed.  As was shown in~\cite{comparin2014}, the surface $X_{W,J_W}$
is polarized by the invariant lattice $S(\sigma_3)=E_6 \oplus U$ and the transcendental lattice
is contained in the lattice $T(\sigma_3) = \, E_8 \oplus A_2 \oplus U \oplus U$. Then
$\rk S(X)\geq 8$.  On the other hand, the surface also admits a  non--symplectic automorphism of order $24$ 
acting by $z \mapsto e^{2i\pi/24} z$
that gives (see Lemma \ref{lem:diag}) $\rk T(X)=8$ or $16$. The second case contradicts   the previous inequality
so that $\rho_X= \rk S(X)=14$.

It  was shown in~\cite{comparin2014} that  the $S(\sigma_3)$-polarized  $X_{W,J_W}$ and the $S(\sigma_3^T)$-polarized surface $X_{W,SL_W}$ 
form a LP mirror pair. Indeed we have:
\begin{itemize}
\item  For $(W,J_W)$, the invariant sublattice is $S(\sigma_3)= E_6 \oplus U$ while $T(\sigma_3)= E_8 \oplus A_2 \oplus U 
\oplus U$. 
\item  For $(W,SL_W)$, the invariant sublattice is $S(\sigma_3^T)=E_8 \oplus A_2 \oplus U$ and $T(\sigma_3^T)=E_6\oplus U\oplus U$.
\end{itemize}

First, for the surface $(W,J_W)$, the action of $\sigma_3$ on $T(\sigma_3)$,  is given as follows (by Remark \ref{unicity} 
the action is unique up to conjugation  by isometries). On the $A_2$ lattice, by taking the lattice metric (Gram matrix)
\begin{equation}\label{eq:j3a2}
A_2^{st}: = \left(\begin{array}{cc} -2& 1\\ 1 & -2 \end{array} \right)\, ,
\end{equation}
one gets
\begin{equation}
M_3^{A_2}:= \left(\begin{array}{cc} 0& -1\\ 1 & -1 \end{array} \right)\, ,
\end{equation}
while on $E_8 \oplus U \oplus U$ 
it is the same as for the previous surface, see equation \eqref{eq:j3uu}. 
Second, for the mirror surface $(W,SL_W)$, the action of $\sigma_3^T$ on $T(\sigma_3^T)$ is given as follows. On $U\oplus U$, it  
is given by the matrix $M_3^{U\oplus U}$ defined in \eqref{eq:j3uu} 
and on $E_6$ the action is given by (see the Appendix)
\begin{eqnarray}\label{eq:j3e6}
M_3^{E_6} :=\left(\begin{array}{cccccc}
0&-1&0&1&0&0\\
1&-1&-1&2&0&0\\
0&0&-2&3&0&0\\
0&0&-1&1&0&0\\
0&0&-1&2&0&-1\\
0&0&-1&1&1&-1
\end{array}
\right)
\end{eqnarray}
in a lattice basis in which the lattice metric is
\begin{eqnarray*}
E_6^{st} :=\left(\begin{array}{cccccc}
-2&1&0&0&0&0\\
1&-2&1&0&0&0\\
0&1&-2&1&1&0\\
0&0&1&-2&0&0\\
0&0&1&0&-2&1\\
0&0&0&0&1&-2
\end{array}
\right).
\end{eqnarray*}

The surface~(\ref{eq:23824}) admits also a non-symplectic automorphism of order 2, acting as 
 $\sigma_2 \, : \ w \mapsto - w$. Following~\cite{Artebani2014758}, the invariant lattice of 
$\sigma_2$ is of rank six and isometric to $S(\sigma_2) \cong 
D_4 \oplus U \subset T(X)$, hence $T(\sigma_2) \cong
E_8 \oplus D_4 \oplus U \oplus U$. The previous theorem indicates 
that the $S(\sigma_2)$-polarized surface $X_{W,J_W}$ and the $S(\sigma_2^T)$-polarized
surface $X_{W,SL_W}$ form a LP mirror pair.  One can check that $(W,SL_W)$ admits an order two  non-symplectic automorphism  
$\sigma_2^T$  of the invariant lattice $S(\sigma_2^T) \cong   E_8 \oplus D_4 \oplus U $.
\end{exa}

\section{Non-geometric automorphisms of K3 sigma-models}
\label{sec:nongeom}

In this section we define {\it mirrored automorphisms} of sigma-model CFTs with K3 target spaces, combining the action of non-symplectic automorphisms of a surface and of its mirror, and study the corresponding isometries of 
the lattice $\Gamma_{4,20}$. This construction is inspired by the 
non-geometric string theory compactifications that were obtained in~\cite{Israel:2013wwa}  as asymmetric orbifolds of Gepner models.

\subsection{Mirrored automorphisms and isometries of the $\Gamma_{4,20}$ lattice}
\label{subsec:diagaut}

We consider a $p$-cyclic K3-surface $X$, associated with a given non-symplectic automorphism $\sigma_p$ of prime order. 
The BH mirror of this surface admits also a non-symplectic automorphism $\sigma_p^T$ of the same order;  
by the theorem~\ref{theo:mir}, these two surfaces, polarized by the invariant sublattices with respect 
to $\sigma_p$ and $\sigma_p^T$ respectively, are also LP mirrors. 
The automorphism $\sigma_p$ has an action on the vector space $T(\sigma_p)^\mathbb{R}$ while the
automorphism $\sigma_p^T$ acts on the vector space $T(\sigma_p^T)^\mathbb{R}$. By the 
corollary~\ref{coro:decomp} these vector spaces are orthogonal to each other in $\Gamma_{4,20}^\mathbb{R}$.

Inspired by physical considerations that will be illustrated in the next subsection, we will consider a mirrored automorphism  
that combines $\sigma_p$ and $\sigma_p^T$ into a non-geometrical automorphism $\hat{\sigma}_p$ of the 
CFT defined by quantizing the non-linear sigma model with a $p$-cyclic K3 surface target space. 
To prove that its action on the lattice $\Gamma_{4,20}$ 
is well-defined we will need the following proposition:

\begin{propo}\label{doubleauto}

Let $X_{W,G}$ be a $p$-cyclic  $S(\sigma_p)$-polarized K3 surface, 
where $S(\sigma_p)$ is the invariant sublattice under the $\sigma_p$ action, with $p$ prime, 
and $X_{W^T,G^T}$ its LP mirror,  regarded as an $S(\sigma_p^T)$-polarized K3 surface.

Observe that $\Gamma_{4,20}$ is an over-lattice of finite index\footnote{A lattice $L$ is an over-lattice  
of a lattice $M$ if $M$ is a sub-lattice of $L$ and if $M$ has finite index $[L:M]$ in $L$ (viewing $M$ as 
a sub-group of the Abelian group $L$), such that both 
lattices have the same rank. The dual lattice $M^\star$ of a lattice $M$ is a free $\mathbb{Z}$-module that 
contains $M$. The quotient $M^\star/M$ is called the discriminant group of the lattice. } 
of the lattice $T(\sigma_p^T) \oplus T(\sigma_p)$,  with index given by $|\det T(\sigma_p)|=|\det T(\sigma_p^T)|$,  
where  for a lattice $L$, $\det L$ is the determinant of the lattice metric.

The equality of the determinants follows from the fact that the lattice $\Gamma_{4,20}$ is unimodular 
(see~\cite[Corollary 2.6]{BPV}). By properties of K3 surfaces 
(see~\cite[Theorem 0.1]{zbMATH03710310} and \cite[Lemma 2.5]{BPV}) the automorphisms $\sigma_p$ and $\sigma_p^T$ act 
trivially on the discriminant groups of $T(\sigma_p^T)$ and of $T(\sigma_p)$ so that we can extend the diagonal action 
by $\sigma_p$ and $\sigma_p^T$ to the whole lattice $\Gamma_{4,20}$.

\end{propo}

Corollary~\ref{coro:decomp} and Proposition~\ref{doubleauto} allow us to define \lq mirrored automorphisms' 
of CFTs with $p$-cyclic K3 surfaces target spaces in the following way: 
\begin{defi}
\label{def:doub}
Let $(X_{W,G},\jmath)$ be a $p$-cyclic K3 surface polarized by 
the invariant sublattice $S(\sigma_p)$ and $(X_{W^T,G^T},\jmath^T)$, polarized by $S(\sigma_p^{\, T})$, 
its LP mirror, with $p\in \{2, 3,5,7,13 \}$.  

By Corollary~\ref{coro:decomp} and Proposition~\ref{doubleauto} the diagonal action by $(\sigma_p,\sigma_p^T)$ on the lattice $T(\sigma_p^T) \oplus T(\sigma_p)$ can be extended to an isometry of the lattice $\Gamma_{4,20}$, that we 
associate with the action of a CFT automorphism denoted $\hat{\sigma}_p$, that we name {\it mirrored automorphism}.

The action of the mirrored automorphism $\hat{\sigma}_p$ on the vector space $\Gamma_{4,20}^\mathbb{R} \cong T(\sigma_p)^\mathbb{R}\oplus T(\sigma_p^T)^\mathbb{R}$, is then given by  
\begin{subequations}
\label{sighat}
\begin{align}
\hat{\sigma}_p^\star \big|_{T(\sigma_p)^\mathbb{R}} & = \sigma_p^\star\\
\hat{\sigma}_p^\star \big|_{T(\sigma_p^T)^\mathbb{R}} & = (\sigma^T_p)^\star
\end{align}
\end{subequations}
\end{defi}
For other values of $p$, including the non-prime cases, 
the physical construction suggests that similar non-geometric automorphisms acting as in eq.~(\ref{sighat}) can 
be defined. However the mathematical classification of these automorphisms is not yet complete (see~\cite{compapriddis}).

The isometry of the lattice $\Gamma_{4,20}$ induced by $\hat{\sigma}_p$ is not 
in the geometric group $O(\Gamma_{3,19}) \ltimes  \mathbb{Z}_{3,19}$ and so is non-geometrical. 
The automorphism $\hat{\sigma}_p$ acts as
\begin{equation}
\hat{\sigma}_p := \mu^{-1} \circ \sigma_p^T \circ \mu  \circ \sigma_p \, ,
\end{equation}
where $\mu$ denotes the BH/LP mirror involution which maps the $K3$ to its mirror; $\mu^{-1}$ maps the mirror $K3$ 
back to the original one, $\sigma_p$ is a diffeomporphism of the original $K3$ and 
$ \sigma_p^T$ is a diffeomorphism of its mirror.

Due to Proposition~\ref{doubleauto} the lattice isometry induced by $\hat{\sigma}_p$ generates the order $p$ isometry subgroup 
\begin{equation}
\label{eq:osigmap}
O(\hat{\sigma}_p) := \langle \hat{\sigma}_p^{\, \star}\rangle \subset O (\Gamma_{4,20})\, .
\end{equation}
Being of finite order, it is conjugate to a subgroup of the maximal compact subgroup 
$$
\big[O(2)\times O(20-\rho_p)\big] \times \big[O(2)\times O(\rho_p)\big] \subset O(2,20-\rho_p)\times 
O(2,\rho_p) 
$$

Explicitly, the action of the non-geometric automorphism $\hat{\sigma}_p$  on $\Gamma_{4,20}$, hence 
on the CFT with a $K3$ target space, is obtained by considering the geometrical action of 
$\sigma_p$ on $T(\sigma_p)$ and, for the BH mirror surface, the action of $\sigma_p^T$ on $T(\sigma_p^T)$. The lattice  
$\Gamma_{4,20}$ is an over-lattice of index $p^k:=|\det T(\sigma_p)|$, $k$ a non-negative integer, 
of the sum $T(\sigma_p) \oplus T(\sigma_p^T)$
(recall that $T(\sigma_p)$ and $T(\sigma_p^T)$ are $p$-elementary lattices i.e. the discriminant groups are sums 
$(\IZ/p\IZ)^{\oplus k}$, $k$ as before).
Then to construct $\Gamma_{4,20}$ one should add to the generators of the lattice $T(\sigma_p) \oplus T(\sigma_p^T)$
exactly $k$ classes of the form $(a+b)/p$ with $a/p$ in the discriminant group of $T(\sigma_p)$ and $b/p$ in the discriminant group $T(\sigma_p^T)$, and we ask also that $((a+b)/k)^2\in\IZ$. 
The action of the isometry $(\sigma_p,\sigma_p^T)$ on these classes is then obtained by linearity (over the rationals). We get in this way a
set of generators of  $\Gamma_{4,20}$ on which we have an isometry $\hat{\sigma}_p$ of order $p$ which is induced by the isometry $(\sigma_p,\sigma_p^T)$ on $T(\sigma_p) \oplus T(\sigma_p^T)$, i.e. the restriction of $\hat{\sigma}_p$ to that lattice is equal to $(\sigma_p,\sigma_p^T)$\footnote{To make the situation more clear we consider a concrete example. This is in fact a more general situation in lattice theory. We consider the hyperbolic lattice $U$ with generators $e$, $f$ such that $e^2=f^2=0$ and $ef=1$. One can primitively 
embed the lattice $\langle 2\rangle \oplus \langle -2\rangle$ into $U$ by sending the two generators to $a:=e+f$ and $b:=e-f$, in this way  the lattice $\langle 2\rangle \oplus \langle -2\rangle$ has index two in $U$. One takes now the isometric involution on $\langle 2\rangle \oplus \langle -2\rangle$ which acts as $\iota:=(\id, -\id)$. The discriminant group of $\langle 2\rangle$,
resp. $\langle -2\rangle$, 
is generated by $a/2$, resp. $b/2$. One considers now the class $w:=(a+b)/2$, which has square $w^2=0\in \IZ$; the lattice generated by $a$ and $(a+b)/2$ has determinant $1$ and it is in fact isometric to $U$. To see this one takes the generators $w$ and 
$v:=(a-(a+b))/2=(a-b)/2$ and the induced involution $\hat\iota$ acts exchanging $v$ and $w$; in particular it cannot be put in a diagonal 
form over $\IZ$.}.

Given that, for any given $p$-cyclic K3 surface, all  the relevant sublattices $T(\sigma_p)$ and $T(\sigma_p^T)$ have been 
tabulated in~\cite{Artebani2014758,comparin2014}, explicit forms of the $\Gamma_{4,20}$ isometries  can be determined 
from lattice theory for any given example, see $e.g.$~\cite{Conway:1987:SLG:39091}. The corresponding matrices  
can be diagonalized on $\mathbb{C}$ according to lemma~\ref{lem:diag} and are characterized respectively by a set of 
$\rk (T(\sigma_p))$ angles and a set of $\rk (T(\sigma_p^T))$ angles; these angles will be discussed further in the following sections.

\begin{exa}
\normalfont

We consider the self-mirror surface~(\ref{eq:self_mirr}) already discussed in example~\ref{exa:selfmirror}. 
The action of the geometrical automorphism $\sigma_3$  on the K3 lattice $\Gamma_{3,19}$ 
has been described there, see eqs.~(\ref{eq:j3uu},\ref{eq:j3e8}). The lattice $\Gamma_{4,20}$ admits 
an orthogonal decomposition into the invariant quantum lattices 
$S^\mathfrak{Q} (\sigma_p)$ and $S^\mathfrak{Q} (\sigma_p^T)$ --~or equivalently into 
$T(\sigma_p^T)$ and $T(\sigma_p)$~-- corresponding 
respectively to the quantum Picard lattice $S^\mathfrak{Q} (X)$ and the transcendental lattice $T(X)$ of 
this surface.

As the surface and its mirror are isomorphic to each other, it is straightforward to define the action of 
the mirrored CFT automorphism $\hat{\sigma}_3$. It has a diagonal action on 
\begin{equation}
\Gamma_{4,20}  \cong \Big( E_8 \oplus U \oplus U\Big) \oplus \Big( E_8 \oplus U \oplus U\Big)\, ,
\end{equation}
duplicating the action of $\sigma_{3}$ on the transcendental lattice that was studied in 
subsection~\ref{subsec:prime}. The action of $\hat{\sigma}_3$ on the sigma-model CFT associated 
with the surface~(\ref{eq:self_mirr}) is therefore given by the block-diagonal $24\times 24$ integer matrix:
\begin{equation}
\label{Mblock}
\hat{M}_3 := \left( \begin{array}{cccc} M_3^{E_8} & 0 & 0 & 0 \\0 & M_3^{U\oplus U} & 0 & 0 \\ 0& 0  & M_3^{E_8}& 0\\
0&0&0&M_3^{U \oplus U} \end{array} \right) \in O(\Gamma_{4,20})\, ,
\end{equation}
where $M_3^{E_8}$ is given by eq.~(\ref{eq:j3e8}) and $M_3^{U\oplus U}$ by eq.~(\ref{eq:j3uu}). 
\end{exa}

\begin{exa}
{\rm 
By considering the K3 surface given by equation~(\ref{eq:23824}) and its mirror we have the orthogonal 
decomposition of the vector space $\Gamma_{4,20}^\mathbb{R}$:
\begin{equation}
\Gamma_{4,20}^\mathbb{R} \cong \Big(U\oplus U\oplus E_6 \Big)^\mathbb{R} 
\oplus \Big(U\oplus U\oplus A_2\oplus E_8\Big)^\mathbb{R} \, .
\end{equation}
The action of $\hat{\sigma}_3$ is then  induced from the block-diagonal $24\times 24$ integer matrix 
\begin{equation}
 \left( \begin{array}{ccccc} M_3^{U\oplus U}&0&0&0&0\\0&M_3^{E_6}&0&0&0\\
0&0&M_3^{U\oplus U}&0&0\\0&0&0&M_3^{A_2}&0\\0&0&0&0&M_3^{E_8}\end{array}\right) \in O\big(T(\sigma_p) \oplus T(\sigma_p^T) \big)\, .
\end{equation}
where the various matrices $M_3^{U\oplus U}$, $M_3^{E_6}$, $M_3^{U\oplus U}$, $M_3^{A_2}$ and $M_3^{E_8}$ are given respectively in eqs.~(\ref{eq:j3uu},\ref{eq:j3e8},\ref{eq:j3a2},\ref{eq:j3e6}). 

This matrix is an isometry of the lattice $T(\sigma_p) \oplus T(\sigma_p^T) \cong U\oplus U\oplus E_6
\oplus U\oplus U\oplus A_2\oplus E_8$. The latter is a sublattice of index $3$ of $\Gamma_{4,20}$  
(see Proposition~\ref{doubleauto})  and more precisely the lattice $E_6\oplus A_2$ is a sublattice of 
index $3$ of $E_8$. Now the isometries of order three $\sigma_3^\star$ and $(\sigma_3^T)^\star$ have 
no fixed vectors on $A_2$, respectively $E_6$ and act trivially on the discriminant groups. They can 
be then combined (see Definition~\ref{def:doub}) to give an isometry of order three on $E_8$ without 
fixed vectors, but up to isometry there is only one such isometry on $E_8$ which is given 
by equation~\eqref{eq:j3e8}. So the action of $\hat{\sigma}_3$ on the sigma-model  CFT
associated with this surface is given by the matrix $\hat{M}_3$ in eq.~(\ref{Mblock}), as for the previous example.
}
\end{exa}

\subsection{Symmetries of Landau-Ginzburg mirror pairs}
\label{subsec:autLG}

The Gepner models arise at special points in the moduli space~(\ref{eq:modsigma}) of sigma-model CFTs on K3 surfaces. 
These Gepner points play a special role in the present context as some of them are fixed under the action of the 
mirrored automorphisms defined in the previous subsection. 

A Gepner model for a K3 surface~\cite{Gepner:1987qi}
is  a $(4,4)$ superconformal field theory obtained as the infrared fixed point of a (2,2) Landau-Ginzburg orbifold~\cite{Vafa:1989xc,Intriligator:1990ua} with Fermat type superpotential
\begin{equation}
\label{eq:LG_sm}
W=Z_1^{k_1} + Z_2^{k_2} + Z_3^{k_3}+ Z_4^{k_4}\, ,
\end{equation}
quotiented by the order $K=\text{gcd}(k_1,\ldots,k_3)$ diagonal $\mathbb{Z}_K$ symmetry $j_W$,  acting on the chiral superfields as 
\begin{equation}
j_W \, : \ Z_\ell \mapsto e^{2i\pi/k_\ell} Z_\ell
\end{equation}
and realizing the projection onto integral R-charges. We will refer to this orbifold 
as the diagonal $\mathbb{Z}_K$ orbifold. The twisted sectors of this orbifold are labelled by 
$\gamma \in \{0,\ldots,K-1\}$ and will be referred to as  $\gamma$-twisted sectors.

The Landau-Ginzburg orbifold/Gepner model with superpotential~(\ref{eq:LG_sm}) has a quantum
Abelian symmetry~\cite{Vafa:1989ih} which is not present in the large-volume limit of the sigma-model. In the diagonal $\mathbb{Z}_K$ orbifold theory, the quantum 
symmetry acts on a field in the $\gamma$-twisted sector of the model as:
\begin{equation}
\sigma^\textsc{q}\, :\ \phi_\gamma \ \mapsto e^{\frac{2i\pi}{K} \gamma} \phi_\gamma\, .
\end{equation}

At the infrared fixed point, the superconformal field theory obtained from this model is an orbifold of a product of 
$\mathcal{N}=2$ minimal model CFTs, as every single-field Landau-Ginzburg model with superpotential 
$W=X^{k_\ell}$ flows to a super-coset CFT $SU(2)_{k_\ell}/U(1)_{k_\ell}$~\cite{Witten:1993jg}. 
These Gepner models lead to IIA superstring theory 
compactifications in six dimensions with $\mathcal{N}=(1,1)$ supersymmetry, or, compactifying further on 
a two-torus, to $\mathcal{N}=4$ supersymmetry in four-dimensions.

\subsubsection*{Asymmetric orbifolds}

We will now explain the relation between the mirrored automorphisms introduced in subsection~\ref{subsec:diagaut} and the non-geometric orbifolds of Gepner models presented in~\cite{Israel:2013wwa,Israel:2015efa}, 
following earlier works~\cite{Schellekens:1989wx,Intriligator:1990ua}.

We consider the Gepner model corresponding to the Landau-Ginzburg orbifold of 
superpotential~(\ref{eq:LG_sm}) and assume that $p:=k_1$ is a prime number. The theory admits the order $p$ symmetry
\begin{equation}
\label{eq:sigma3}
\sigma_p\, : \ Z_1  \mapsto e^{2i\pi/p} Z_1\, .
\end{equation}
Quotienting the Gepner model by this automorphism alone would break all space-time supersymmetry. 
Indeed one can see that in the corresponding orbifold theory (see~\cite{Israel:2013wwa} for details):
\begin{itemize}
\item all the worldsheet operators corresponding to space-time supercharges are charged 
under the $\mathbb{Z}_p$  symmetry hence are projected out of the spectrum,
\item the $b$-twisted sectors of the $\mathbb{Z}_p$ orbifold by the symmetry~(\ref{eq:sigma3}) contain states 
with non-integer left and right $U(1)_R$-charges whenever $b\neq 0$.
\end{itemize}
One observes that one can define a subgroup of the quantum symmetry group of the model~(\ref{eq:LG_sm}), isomorphic to $\mathbb{Z}_p$, 
generated by:
\begin{equation}
\sigma^{\textsc{q}}_p := (\sigma^\textsc{q})^{K/p} \, .
\end{equation}

One can then modify the orbifold of the LG orbifold/Gepner model~(\ref{eq:LG_sm}) 
by the symmetry~(\ref{eq:sigma3}) that we described above by adding a specific {\it discrete torsion} 
keeping the space-time supercharges coming from the left-moving sector in the spectrum. 

Starting from the Gepner model, one defines the $\mathbb{Z}_p$ orbifold projection by assigning to every state in the theory a charge 
\begin{equation}
\label{eq:q3_torsion}
\hat{Q}_p \equiv  Q_p + Q_p^\textsc{q}  \mod 1  \equiv Q_p + \frac{\gamma}{p} \mod 1 \, ,
\end{equation}
where $Q_p$ 
is the charge of the given state under the action of $\sigma_p$ and $Q_p^\textsc{q}$ is the charge 
under the quantum symmetry $\sigma^{\textsc{q}}_3$, and by projecting onto states with $\hat{Q}_p \in \mathbb{Z}$. 
This discrete torsion has also an 
effect in the twisted sectors $b \neq 0$ of the new $\mathbb{Z}_p$ orbifold. 
In those sectors the diagonal $\mathbb{Z}_K$ orbifold projection is modified, 
as one projects onto states with $\hat{Q}_K \in \mathbb{Z}$, where
\begin{equation}
\label{eq:qk_torsion}
\hat{Q}_K \equiv  Q_K - \frac{b}{p}\mod 1\, .
\end{equation}
The charge assignments~(\ref{eq:q3_torsion}) and~(\ref{eq:qk_torsion}) are related to each other by modular invariance.  

One can check, by inspecting the one-loop partition function, that the $\mathbb{Z}_p$ orbifold projection w.r.t. 
the charge $\hat{Q}_p$  keeps all space-time supercharges from the left-movers, while none 
of the space-time supercharges from the right-movers is invariant. Furthermore the diagonal $\mathbb{Z}_K$ projection w.r.t. the charge $\hat{Q}_K$ keeps only states with integer left R-charge. Hence space-time supersymmetry 
from the left-movers on the worldsheet is preserved by this orbifold with discrete torsion. Notice that one could have used the charge $Q_p - Q_p^\textsc{q}$ instead, in which case the invariant 
space-time supercharges come from the right-movers.

Under mirror symmetry, the right R-charges in every $\mathcal{N}=2$ minimal  model are mapped to minus themselves. As a consequence, mirror symmetry exchanges the geometrical automorphism 
$\sigma_p$  with its quantum counterpart $\sigma^{\textsc{q}}_p$. In view of the discussion in section~\ref{sec:math}, a generator $\sigma^{\textsc{q}}_p$ of an order $p$ subgroup 
of the quantum symmetry of a Landau-Ginzburg orbifold superconformal field theory 
is identified with a non-symplectic automorphism $\sigma_p^T$ of the corresponding BH mirror K3 surface. Hence, 
the mirrored   automorphisms introduced in subsection~\ref{subsec:diagaut} correspond precisely, 
at the Gepner points in the moduli space, to the orbifolds with discrete torsion described here.

The two-fold choice of discrete torsion in the definition of the Landau-Ginzburg model 
symmetries $\hat{Q}_p:= Q_p \pm Q_p^\textsc{q}$  that we have noticed above corresponds, in the 
language of subsection~\ref{subsec:diagaut}, to the possibility of pairing the action of $\sigma_p$ either with the 
action of $\sigma_p^T$ or of its inverse.

\subsubsection*{Fractional mirror symmetry}

We have described in the previous subsection orbifolds of Gepner models with discrete torsion, that preserve all 
space-time supersymmetry from the left-movers, and none from the right-movers at first sight.  
As discussed in~\cite{Israel:2015efa}, they belong to a more general family of quotients of Gepner models 
by non-symplectic automorphisms of the corresponding K3 surfaces with discrete torsion. This construction leads 
generically to non-geometric compactifications, $i.e.$ that do not belong to the moduli space of compactifications on smooth manifolds, preserving $\mathcal{N}=2$ supersymmetry in four dimensions (after further compactification on $T^2$). Similar constructions exist for 
Calabi-Yau three-folds, leading to $\mathcal{N}=1$ in four dimensions.

However, in the specific case of an orbifold of a $p$-cyclic K3 surface (or more generically of a 
$p$-cyclic CY manifold) by a non-symplectic automorphism of order $p$ as considered in the present work, 
the twisted sectors $b\neq 0$ contain right-moving operators that, despite having non-integral right 
$R$-charge, have the properties of generators of space-time supersymmetry. 
Interestingly, the worldsheet model for the non-geometric compactification is actually isomorphic as a 
$(2,2)$ superconformal field theory to the original Calabi-Yau model in this case.

This isomorphism implies the existence of quantum symmetries, called {\it fractional mirror symmetries} in~\cite{Israel:2015efa}, 
between geometric and non-geometric compactifications in the Landau-Ginzburg regime. Outside of the Gepner point such symmetry 
extends to a map between a $(2,2)$ non-linear sigma-model on a Calabi-Yau manifold and a non-geometric worldsheet model. 
A linear-sigma model description~\cite{Witten:1993yc} was proposed in~\cite{Israel:2015efa}, 
generalizing the ideas of~\cite{Hori:2000kt}.

In the following, we will focus on freely-acting orbifolds combining this type of $K3$ non-geometrical orbifold with a shift 
along a circle; in this case the accidental isomorphism and corresponding restoration of $\mathcal{N}=4$ space-time 
supersymmetry do not play a role, as the $b$-twisted sectors with $b\neq 0$ will only contain massive states 
from the space-time point of view.

\subsection{Worldsheet construction of non-geometric backgrounds: summary}
\label{subsec:free}
The mirrored automorphisms described in subsections~\ref{subsec:diagaut} for the geometry and~\ref{subsec:autLG} for 
the field theory are the building blocks of non-geometric compactifications of type IIA superstring theory, 
whose vacua correspond to the non-geometric freely-acting orbifolds of~\cite{Israel:2013wwa} that we will 
now summarize briefly.

The starting point is a Gepner model for a K3 surface as described in subsection~\ref{subsec:autLG}. 
Consider the tensor product of this superconformal field theory with the free $c=3$ superconformal theory with a 
two-torus target space of coordinates $Y_1$, $Y_2$. We consider a freely-acting 
supersymmetry-breaking $\mathbb{Z}_{k_1} \times \mathbb{Z}_{k_2}$ orbifold of this $K3\times T^2$ 
superconformal model generated by
\begin{subequations}
\begin{align}
g_1 &: \ Z_1 \mapsto e^{2i\pi/k_1} Z_1 \ , \ Y_1 \mapsto Y_1+ 2\pi/k_1\\
g_2 &: \ Z_2 \mapsto e^{2i\pi/k_1} Z_2 \ , \ Y_2 \mapsto Y_2+ 2\pi/k_2
\end{align}
\end{subequations}
As it is, this orbifold breaks all space-time supersymmetry for the reasons given in the previous subsection.

We add to each of these two freely-acting orbifold actions a {\it discrete torsion} of the same type as in 
eqs.~(\ref{eq:q3_torsion},\ref{eq:qk_torsion}) above. As described there the discrete torsion is such that all these 
models have integral left-moving R-charges but non-integral right-moving ones, hence a type IIA superstring theory 
built upon one of these models will have a four-dimensional Minkowski vacuum with $\mathcal{N}=2$ space-time 
supersymmetry, all space-time supercharges being obtained from the left-moving Ramond ground states. 

The two gravitini obtained from the right-moving Ramond sector are indeed massive in the $\mathbb{Z}_{k_1} \times \mathbb{Z}_{k_2}$ 
orbifold theory. Because of the freely-acting nature of this orbifold, 
no massless states could possibly arise from the corresponding 
twisted sectors. If one chooses an orthogonal two-torus of radii $R_1$ and $R_2$, the masses squared of the two massive 
gravitini of broken $\mathcal{N}=4$ supersymmetry are~\cite{Israel:2013wwa}:
\begin{equation}
m^2 = \left(\frac{1}{k_1 R_1}\right)^2 + \left(\frac{1}{k_2 R_2}\right)^2\, .
\end{equation}

The massless spectra of all these models were computed in~\cite{Israel:2015efa}.\footnote{In~\cite{Israel:2015efa} all 
pair of purely non-symplectic automorphisms were considered, for prime and non-prime order $p$.} 
It was found there that the massless states are identified with a subset of the chiral rings of 
the $K3$ SCFT, containing states built out of the identity operator in the $SU(2)_{k_1}/U(1)_{k_1}$ and 
$SU(2)_{k_2}/U(1)_{k_2}$ minimal models; we will consider the corresponding moduli spaces in subsection~\ref{subsec:mod_space} from another perspective. 

In about half of the possible constructions, this subset is empty and hence all the moduli of 
the original $K3$ SCFT have become massive; the only remaining massless moduli are the $T$ and $U$ moduli of the two-torus and the 
axio-dilaton modulus $S$, that are part of space-time vector multiplets. It gives the $\mathcal{N}=2$ four-dimensional $STU$ 
supergravity model at low energies (compared to the inverse size of the torus). In the remaining constructions, some of the $K3$ 
moduli survive and appear in the low energy theory in massless hypermultiplets. We now turn to the second part of this article, 
where we analyse these constructions from the low-energy four-dimensional viewpoint. 


\section{ $\mathcal{N}=4$ gauged supergravity from duality twists}
\label{sec:gauge}

In this section we study the supergravity dimensional reduction that corresponds to  the stringy  construction considered in 
the previous sections. We have considered type IIA superstring theory compactified on $K3 
\times T^2$ identified under  certain automorphisms. This requires being at a point in the $K3$ moduli space which is a fixed point under 
the automorphisms. (These fixed points were found in the last section from Landau-Ginzburg orbifolds.) This construction  is extended 
to general points in moduli space by a compactification with duality twists~\cite{Dabholkar:2002sy}. We will here discuss the 
supergravity limit of this, which is a dimensional reduction of Scherk-Schwarz type~\cite{Scherk:1979zr}.

We consider then type IIA superstring theory compactified on $K3 
$ to 6 dimensions and then further compactified on $ T^2$ with duality twists with non-geometric monodromy. 
In the supergravity limit, compactifying  IIA supergravity on $K3$ gives $\mathcal{N}=(1,1)$ supergravity 
in six dimensions coupled to 20 vector multiplets, and this has a duality symmetry $O(4,20)\times \mathbb{R}$. Then further 
compactifying on $T^2$ with an $O(4,20)$ monodromy round each circle gives a Scherk-Schwarz reduction of the supergravity, resulting in a gauged $\mathcal{N}=4$ supergravity in four dimensions. 
This construction has been discussed  extensively  in the supergravity literature; see e.g.~\cite{Scherk:1979zr,Hull:2005hk,Kaloper:1999yr,Porrati:1989jk}
and references therein. For our  string theory constructions,  the monodromies are required to be in the duality group 
$O(\Gamma_{4,20})$, $i.e.$ the isometry group of the lattice of total cohomology of the $K3$ surface
as was discussed in section~\ref{sec:math}; in the physics literature it is often refered to as $O(4,20;\mathbb{Z})$.

We will focus here on  the case with monodromies  that are in the $O(4)\times O(20)$ subgroup of $O(4,20)$ 
as it is  for these compact monodromies that fixed points in the moduli space corresponding to Minkowski minima of 
the $D=4$ supergravity scalar potential are possible.
As we shall see, some interesting features arise for these special cases, and will give some vacua that break the 
$\mathcal{N}=4$ supersymmetry in four dimensions to $\mathcal{N}=2$. 
We will summarize the supergravity results here, and give more details elsewhere.

In this section we will consider the supergravity reduction with monodromies in the continuous group $O(4,20)$ and 
in the following section we will consider the consistent type IIA superstring theory compactifications 
that arise from the discrete monodromies constructed in section~\ref{sec:nongeom}.


\subsection{Twisted reduction on $T^2$}
\label{subsec:twisted_red}
The starting point in six dimensions  is $\mathcal{N}=(1,1)$ 
supergravity coupled to 20 vector multiplets, and this has a rigid duality symmetry $G=O(4,20)$ 
(and a further rigid symmetry consisting of constant shifts of the dilaton).  
There is also a local symmetry   which  in the bosonic sector is $H= O(4)\times O(20)$.
In extending to the fermionic sector, the local symmetry is actually a double cover of this,
$H_s= Pin(4)\times O(20)$.
The 24 vector fields $A^I_m$ ($I=1,\dots , 24$) transform as the {\bf 24} of $G=O(4,20)$ and are invariant under $H$.
The fermions are invariant under $G=O(4,20)$ but transform under $H_s$. The scalars consist of a dilaton $\phi$ and 
scalars taking values in the coset $G/H$ and can be represented by a $G$-valued vielbein field
 $\hat {\cal{V}}$
  transforming as $\hat{\cal{V}}\to g \hat{\cal{V}}h^{-1}$ under the action of $h\in H,g\in G$. The vielbein $\hat {\cal{V}}$
 represents $\rm{dim} (G)= 276+1$ degrees of freedom, 
but $\rm{dim} (H)= 196$ of these can be removed by local $H$ transformations.

We now turn to dimensional reduction on $T^2$ with twists in $G$, giving rise to a gauged supergravity in four dimensions. 
Consider first the untwisted case. Simple dimensional reduction (with no twists) on $T^2$ gives rise to 
$\mathcal{N}=4$  supergravity coupled to 22 vector multiplets in four dimensions. The massless Abelian theory 
has $G=SL(2)\times O(6,22)$ global symmetry, and a local symmetry $Pin (6)\times O(22)$, acting on the 
bosonic sector through $O(6)\times O(22)$. 
The $\mathcal{N}=4$ supergravity multiplet contains the vielbein, four gravitini $\psi^i_\mu$, six graviphotons $A^m_\mu$, four 
spin-half fermions $\chi^i$ and a complex scalar $\tau$, which takes value in $SL(2)/SO(2)$. The $SL(2)$ acts 
as usual through fractional linear transformations $\tau \mapsto (a\tau + b)/(c\tau + d)$. 

The 22 vector multiplets in four dimensions each contain  a vector $A^a$, four gaugini $\lambda^{ai}$ and six real scalars.  132 scalars parameterize the coset space $O(6,22)/O(6)\times O(22)$
while the remaining two parameterize the coset space $SL(2)/U(1)$. The scalars in $O(6,22)/O(6)\times O(22)$ can be 
conveniently expressed in terms of a vielbein $\mathcal{V}  \in O(6,22)$, such that 
they transform under global $G=O(6,22)$ and local 
$O(6)\times O(22)$ as $\mathcal{V} \mapsto h^{-1} \mathcal{V} g$. 
From the vielbein $\mathcal{V}$, one can construct a metric $\mathcal{M}$ on the coset space given by
$\mathcal{M}=\mathcal{V}^t \mathcal{V}$ which is invariant under $H$ and transforms tensorially 
under $G$: $\mathcal{M} \mapsto g^t \mathcal{M} g$. 
The group $O(6,22)$ preserves a metric $\eta_{MN}$  of signature $(6,22)$ and for the supergravity theory we can 
choose a basis in which this is the diagonal metric $ ( {\mathbb{I}}_6, -{\mathbb{I}}_{22})$.  However, in the 
next section when we apply the supergravity analysis to string theory, we will take $\eta_{MN}$ to be a lattice metric 
on $\Gamma_{4,20}\oplus U \oplus U$.

A gauged version of this supergravity (with electric gauge group) can be obtained by choosing a subgroup $K$ of the 
rigid $G=O(6,22)$ symmetry (of dimension 28 at most) and promoting 
it to a local symmetry,  using a minimal coupling to the 28 vector fields already in the theory.\footnote{Note that this is 
not the most general gauging of the supergravity, but we will restrict ourselves to this class of gaugings here.} 
For this to work, the vector representation of $O(6,22) $ must be the adjoint representation of $K\subset O(6,22) $.
The gauging of the four-dimensional supergravity is completely specified by the structure constants $t_{MN}{}^P$ 
of the gauge group $K$ ($M,N=1,\dots 28$) which satisfy the Jacobi identity and the constraint that $K$ preserves 
the $O(6,22)$ invariant metric $\eta_{MN}$ which is the condition   that 
$$
t_{MNP}=\eta_{MQ}t_{NP}{}^Q
$$
is completely antisymmetric.
Supersymmetry then requires the addition of a scalar potential $V$, together with fermion mass terms
 given by 
\begin{equation}
 \label{gravmass}
e^{-1} \mathcal{L}_{3/2} = \tfrac{1}{3} A_1^{ij} \bar \psi_{\mu i}
 {\Gamma}^{\mu \nu} 
 \psi_{\nu j} +  \tfrac{1}{3} A_2^{ij} 
 \bar \psi_{\mu i}
  \Gamma ^\mu   \chi_j - A_{2ai}{}^{ j} 
  \bar \psi_{\mu }^i
\Gamma ^\mu \lambda^{a}_j + h.c.
\end{equation}
and 
\begin{equation}
\label{fermmass}
e^{-1} \mathcal{L}_{1/2} =- A_{2ai}^{\ \ \ j} 
\bar \chi^i (\lambda^{a})_j   + \tfrac{1}{2} A_2^{ij} \bar \lambda^{a}_i  \lambda_{aj} + 
A_{ab}^{\ \ ij} 
 \bar \lambda^{a}_i  \lambda_{j}^b
 + h.c. 
\end{equation}
in terms of certain scalar-dependent tensors
$A_1^{ij} , A_2^{ij} , A_{2ai}^{\ \ j}, A_{ab}^{\ \ ij}$.
Here $i,j=1,\dots 4$ are $SU(4)$ indices.
Supersymmetry and gauge invariance put strong restrictions on the subgroups $K$ that can be gauged, 
and fixes the form of the scalar potential and the tensors
$A_1^{ij} , A_2^{ij} , A_{2ai}^{\ \ j}, A_{ab}^{\ \ ij}$; see~\cite{Weidner:2006rp,Schon:2006kz,deWit:2007kvg}
and references therein.
In particular, the scalar potential is
\begin{equation}\label{potential}
V=\frac{1}{48 \Im(\tau)}t_{MNP}t_{QRS}\left({\cal M}^{MQ}{\cal M}^{NR}{\cal M}^{PS}-3{\cal M}^{MQ}\eta^{NR}\eta^{PS}\right)
+\frac{1}{24\Im(\tau)}t_{MNP}t^{MNP}
\end{equation}
where   $\Im(\tau)$ denotes the  imaginary part of $\tau$, the scalar in $SL(2)/U(1)$ and ${\cal M}^{MQ}$ 
is the metric on $O(6,22)/O(6)\times O(22)$ discussed above.

We now turn to the twisted reduction of the six-dimensional supergravity on $T^2$ to obtain a gauged 
$\mathcal{N}=4$ supergravity. We consider a rectangular torus for simplicity, with
coordinates  
$y^i$, $i=1,2$, with $y^1\sim y^1+ 2\pi R_1$ and $y^2\sim y^2+ 2\pi R_2$. Then the complex K\"ahler modulus is
$T=iR_1 R_2$ and complex structure modulus is $U = i R_2/R_1$.

We introduce twists around each of the two circles as described in the introduction.
A field $\psi (x^\mu, y^i)$ (where $y^i$ are coordinates on $T^2$ and   $x^\mu$, $\mu =0, \dots , 3$, are 
the coordinates of the four-dimensional space-time)  
is taken to depend on $y$ through $G$ transformations.
Specifically, suppose $\psi (x^\mu, y^i)$ transforms in a representation of $G$, 
$\psi \mapsto R[g]  \psi
$ under a rigid transformation $h\in H_R$. Then the Scherk-Schwarz ansatz is
\begin{equation}
\psi (x^\mu, y^i)= R[g_1(y^1) ]\,  R[g_2(y^2) ]\, \psi_0 (x) \, ,
\end{equation}
giving the $D=6$ field $\psi (x^\mu, y^i)$ as the transformation of a $D=4$ field $ \psi_0 (x)$ under a $y$-dependent $G$ transformation $g_1(y^1)$ around the first cycle  and a $y$-dependent $G$ transformation
$g_2(y^2)$ around the second cycle.   The two $G$ transformations are required to commute, $$g_1(y^1)g_2(y^2)=g_2(y^2)g_1(y^1)\, ,$$
 and the $y$-dependence is taken to be exponential, so that
\begin{equation}
\label{SchSch}
g_1(y^1)  = e^{N_{1}  y^1} \ , \quad 
g_2(y^2)  = e^{N_{2}  y^2} \, .
\end{equation}
Then the monodromies are
\begin{equation}
\label{eq:monod}
(g_1(0))^{-1}g_1(2\pi R_1)= e^{2\pi R_1 \, N_{1} }, \qquad
(g_2(0))^{-1}g_2(2\pi R_2)= e^{2\pi R_2 \, N_{2} }
\end{equation}
 for two commuting elements $N_1,N_2$ of the Lie algebra of $G$, $[N_1,N_2]=0$.
 In the six-dimensional supergravity, the only fields transforming under $G$ are the vector fields $A$ and the scalar 
fields, represented by the vielbein $\hat{{\cal V}}$. These then get non-trivial $y$ dependence, while the fermions 
and graviton do not, as they are singlets under $G$. This picture depends on using the formalism in which the local 
$H$ symmetry is not fixed. Choosing a physical gauge for the local $H$ symmetry would mean that  a $G$ 
transformation must be accompanied by a compensating $H$ transformation that act on the fermions through an 
$H_s$ transformation, so that in this gauge the fermions also get $y$ dependence, and this requires the choice of 
a lift of the twist in $H$ to one in the double cover $H_s$.
 
Full details of the reduction for the bosonic sector are given in~\cite{ReidEdwards:2008rd}, and here we will just quote the results 
needed, mostly following the notation of~\cite{ReidEdwards:2008rd}.
The $O(4,20)$ invariant metric is $\eta _{IJ}$ where $I,J=1,\dots 24$; for the supergravity theory we can choose a basis in which this is the diagonal metric $ ( {\mathbb{I}}_4, -{\mathbb{I}}_{20})$.  However, in the next section when we apply the supergravity analysis to string theory, we will take $\eta_{IJ}$ to be the metric on the lattice $\Gamma_{4,20}$.
The  generators of the gauge group $K$ can be combined into a $O(6,22)$ vector $T_M$ as
\begin{eqnarray}\label{generator decomposition 2}
T_M=\left(%
\begin{array}{c}
  Z_i \\
  X^i \\
  T_I \\
\end{array}%
\right) \, .
\end{eqnarray}
The Lie algebra of $K$ is then~\cite{ReidEdwards:2008rd}
$$[T_M,T_N]=t_{MN}{}^PT_P \, , $$
 where the structure constants of the gauge group are
\begin{equation}
\label{eq:structconst_def}
t_{iI}{}^J=N_{iI}{}^J,   \qquad  N_{IJi}=\eta_{JK}f_{iI}{}^K,
\end{equation}
and all other structure constants are zero. Then the only non-vanishing commutators are
\begin{equation}
[Z_i , T_I]=N_{iI}{}^J T_J, \qquad [T_I,T_J]= N_{IJi} X^i \, .
\end{equation}

 Suppose now that there is a point $\hat {\cal {V}}=\hat {\cal {V}}_0$ in the moduli space $O(4,20)/O(4)\times O(20)$ that is fixed under both the monodromies. 
 From the arguments of~\cite{Dabholkar:2002sy},  the fixed point gives a minimum of the scalar potential where the potential vanishes, giving a Minkowski vacuum.
 Then we can perform an $O(4,20)$ transformation 
 with $g=(\hat {\cal {V}}_0)^{-1}$ to bring the fixed point to the origin,
 $\hat {\cal {V}}_0=1$. The subgroup of  $O(4,20)$ preserving $\hat {\cal {V}}_0=1$ is $O(4)\times O(20)$ so both 
monodromies (\ref{eq:monod}) must be in this subgroup. 

The supergravity reduction is then specified by two commuting monodromies $N_1$ and $N_2$ in the Lie algebra of
$O(4)\times O(20)$. These are then in a Cartan subalgebra $SO(2)^{12}$ and we can choose a basis (by $O(4)\times O(20)$ 
conjugation) in which they are both diagonal and each specified by $12$ angles:
\begin{align}
\label{angless}
 2\pi R_1N_1 = & \begin{pmatrix} 0&\theta _1\\-\theta _1&0 \end{pmatrix}
 \otimes
  \begin{pmatrix} 0&\theta _2\\-\theta _2&0 \end{pmatrix}
  \otimes \dots 
  \otimes
   \begin{pmatrix} 0&\theta _{12}\\-\theta _{12}&0 \end{pmatrix}\, ,
\notag\\
 2\pi  R_2 N_2 =& \begin{pmatrix} 0&\tilde \theta _1\\-\tilde \theta _1&0 \end{pmatrix}
 \otimes
  \begin{pmatrix} 0&\tilde\theta _2\\-\tilde\theta _2&0 \end{pmatrix}
  \otimes \dots 
  \otimes
   \begin{pmatrix} 0&\tilde\theta _{12}\\-\tilde\theta _{12}&0 \end{pmatrix}\, .
\end{align}
The monodromies in $O(4)\times O(20)$  are then
\begin{align}
\label{monods}
e^{2\pi R_1 \, N_{1} }= & \begin{pmatrix} \cos {\theta _1}&\sin \theta _1\\-\sin \theta _1& \cos {\theta _1} \end{pmatrix}
   \otimes \dots 
  \otimes
  \begin{pmatrix} \cos {\theta _{12}}&\sin \theta _{12}\\-\sin \theta _{12}& \cos {\theta _{12}} \end{pmatrix}\, ,
\notag\\
e^{2\pi R_2 \, N_{2} } =
 & \begin{pmatrix} \cos {\tilde\theta _1}&\sin \tilde \theta _1\\-\sin \tilde\theta _1& \cos {\tilde \theta _1} \end{pmatrix}
   \otimes \dots 
  \otimes
  \begin{pmatrix} \cos {\tilde\theta _{12}}&\sin \tilde \theta _{12}\\-\sin \tilde\theta _{12}& \cos {\tilde\theta _{12}} \end{pmatrix}\, .
\end{align}
We choose a basis in which the angles $\theta _1,\theta _2$ and $\tilde\theta _1, \tilde\theta _2$ specify 
monodromies in the $O(4)$ factor and the remaining angles specify monodromies in $O(20)$.

These twists will   in general give masses to fields that are charged under $U(1)^{12}\subset O(4)\times O(20)$.
For a state with charges $q_i
$ $(i,j=1,\dots 12$)  under $U(1)^{12}$, 
the mass $m$ will be given by
\begin{equation}
\label{masses}
m^2 = 
\left(\sum _{i=1}^{12}
\frac {q_i \theta_i }{ 2\pi R_1}
\right) ^2 + 
\left(\sum _{i=1}^{12}
\frac {q_i  \tilde \theta_i }{  2\pi R_2}
\right)^2
\, .
 \end{equation}
 Using this formula, the masses of all fields can be found by finding the charges $q_i$. 

The 28 vector fields are in the {\bf 28} of  $O(6,22)$ and this decomposes into 
$(\mathbf{4},\mathbf{1})+ (\mathbf{1},\mathbf{24})$ under $O(2,2)\times O(4,20)$.
A twist with all angles non-zero makes the  vectors in the $(\mathbf{1},\mathbf{24})$ representation massive and leaves 
the $(\mathbf{4},\mathbf{1})$ vectors massless. The 24 vector fields in the 
$(\mathbf{1},\mathbf{24})$ representation can be written as
12 complex vector fields $A_i$, where $A_i$ has charge $q_i=1$ and $q_j=0$ for $j\ne i$.  
Then $A_i$ has mass $m$ given by
\begin{equation}
\label{massesv}
m^2 = 
\left( \frac {q_i \theta_i }{ 2\pi R_1}
\right) ^2 + 
\left(
\frac {q_i  \tilde \theta_i }{  2\pi R_2}
\right)^2\, .
 \end{equation}
 If  the angles  $\theta _i, \tilde \theta _i$ are both zero for some $i$, then the vector   $A_i$ will remain massless.

The scalars in $O(6,22)/O(6)\times O(22)$ can be parameterized by fields
transforming as the $(\mathbf{6},\mathbf{22})$ under $O(6)\times O(22)\subset O(6,22)$.
Under $O(2)\times O(2)\times O(4)\times O(20)$
 these decompose as\footnote{Here
$O(2)\times O(4)\times \subset O(6)$ and $O(2)\times O(20)\subset O(22)$.} 
\begin{equation}
(\mathbf{2},\mathbf{2},\mathbf{1},\mathbf{1})+(\mathbf{2},\mathbf{1},\mathbf{1},\mathbf{20})+(\mathbf{1},\mathbf{2},\mathbf{4},\mathbf{1})+(\mathbf{1},\mathbf{1},\mathbf{4},\mathbf{20})\, .
 \end{equation}
Of these, only those in the $(\mathbf{2},\mathbf{2},\mathbf{1},\mathbf{1})$ 
representation are singlets under $O(4)\times O(20)$ and hence invariant under the $U(1)^{12}$ twist.
These scalars parameterize
$$\frac {O(2,2)}{O(2)\times O(2)}
 \subset 
\frac {O(6,22)}{O(4)\times O(22)} \, .
$$
The axion and dilaton in $SL(2)/U(1)$ are also uncharged and remain  massless, so the
scalars in
$$
\frac {SL(2)}{U(1) }
\times \frac { O(2,2)}{O(2)\times O(2) }
\sim \left[
\frac {SL(2)}{U(1) }
\right]^3
$$
remain massless.

All other scalars become generically massive, with masses given by eq~(\ref{masses}). This formula indicates also that, for 
a given set of charges $\{ q_i, i=1,\ldots,12\}$,  some values of the angles $\theta _i$, $\tilde \theta _i$ can lead to 
accidentally massless scalars.

It will be useful to decompose the indices $i=1,\dots 12=(M,A)$ into indices $M=1,2$ 
labeling the Cartan subalgebra of $O(4)$ 
and indices  $A = 3,\dots 12$ labeling the Cartan subalgebra of $O(20)$, so that the charges are $q_i=(q_M,q_A)$.
 Then, for example, the 80 real scalars in the $(\mathbf{1},\mathbf{1},\mathbf{4},\mathbf{20})$ representation take values in 
the coset $O(4,20)/O(4)\times O(20)$ and can be written in terms of complex scalars $\phi _{MA}, \rho_{MA}$
where the
scalar $\phi _{NB}$ has charges $q_i=(q_M,q_A)$ where $q_M = \delta _{MN}$ and 
$q_A = \delta _{AB}$ while $\rho _{NB}$ has charges $q_i=(q_M,q_A)$ where $q_M = \delta _{MN}$ and 
$q_A =- \delta _{AB}$.
Then 
$\phi _{MA}$ has mass squared
\begin{equation}
\label{massess}
m^2 = 
\left( 
\frac {  \theta_M+ \theta_A }{ 2\pi R_1}
\right) ^2 + 
\left(\ 
\frac {\tilde  \theta_M+ \tilde \theta_A}{  2\pi R_2}
\right)^2 \, ,
 \end{equation}
and $\rho_{MA}$
has mass squared
\begin{equation}
\label{massesss}
m^2 = 
\left( 
\frac {  \theta_M- \theta_A }{ 2\pi R_1}
\right) ^2 + 
\left(\ 
\frac {\tilde  \theta_M- \tilde \theta_A}{  2\pi R_2}
\right)^2 \, . 
\end{equation}

Then for generic twists with all angles non-zero, the massless bosonic fields consist of 
the graviton, 4 vector fields in the $\mathbf{4}$ of $O(2,2)$ and 6 scalars in the coset space $[SL(2)/U(1)]^3$; 
this is precisely the bosonic sector of the STU model~\cite{Duff:1995sm}.

We now turn to the fermion mass terms (\ref{gravmass}),(\ref{fermmass}). At the origin,  $\mathcal{V} = \mathbb{I}_{28}$, 
 the mass matrices of the model simplify considerably to give~\cite{Horst:2012ub}
\begin{align}
\label{eq:cst_matrices}
&A_1^{ij} = A_2^{ij} = \tfrac{1}{8\sqrt{\tau_2}} ([G_m]_{ik})^\star [G_n]_{kl} ([G_p]_{lj})^\star t_{mnp} \ , 
\quad A_{2ai}^{\ \ \ j} = -\tfrac{1}{4\sqrt{\tau_2}} [G_m]_{ik} ([G_n]_{kj})^\star t_{amn} \ , \notag \\ 
&A_{ab}^{\ \ \ ij} = - \tfrac{1}{2\sqrt{\tau_2}} [G_m]^{ij} t_{abm}
\end{align}
where $G_m$ are the 't Hooft matrices used to convert an $SO(6)$ vector index to an antisymmetric 
pair of $Spin(6)=SU(4)$ indices.
The first matrix $A_1$ gives direct access to the fraction of supersymmetry preserved by the vacuum, as it 
provides the mass term for the gravitini given by 
\begin{equation}
 \tfrac{1}{3} A_1^{ij} \bar \psi_{\mu i}
 {\Gamma}^{\mu \nu} 
 \psi_{\nu j}+ h.c.
\end{equation}
where $A_1^{ij} $ is a complex symmetric matrix.
The mass matrix for $\psi_{\mu i}$ is
\begin{equation}
(M^2)_i{}^j
= A_{1ik} A_1^{kj} 
\end{equation}
where $A_{1ij}= (A_1^{ij} )^*$.
This is a Hermitian matrix whose eigenvalues are (after a calculation using formul\ae{}  from~\cite{Horst:2012ub}) $(m_1)^2$ and $(m_2)^2$, both with degeneracy two, where
 \begin{equation}
 \label{fermmas1}
 (m_1)^2 = 
 \left(
\frac {\theta_1 -\theta _2}{ 4\pi R_1}\right) ^2+ \left(
\frac {\tilde\theta_1 - \tilde\theta _2}{ 4\pi R_2}
\right) ^2 
 \end{equation}
and
\begin{equation}
 \label{fermmas2}
 (m_2)^2 = 
 \left(
\frac {\theta_1 + \theta _2}{ 4\pi R_1}\right) ^2
+ \left(
\frac {\tilde\theta_1 + \tilde\theta _2}{ 4\pi R_2}
\right) ^2 \, .
\end{equation}

These formul\ae{}  can be understood as follows. The  $D=6$ supergravity has a local $H_s=Pin(4)\times O(20)$ 
symmetry and a global $O(4,20)$ symmetry. The fermions transform under $Spin(4)\times O(20)=SU(2)
\times SU(2)\times O(20)\subset H_s$, but do not transform under the  global $O(4,20)$ symmetry. 
If the local $H_s$ symmetry is fixed,  $O(4,20)$ transformations must be accompanied by compensating $H_s$ transformations.
As a result, reductions with $O(4,20)$  twists result in twists of the fermions by compensating $H_s$ transformations.
The gravitini transform as  $(\mathbf{2},\mathbf{1},\mathbf{1})+(\mathbf{1},\mathbf{2},\mathbf{1})$
under $SU(2)
\times SU(2)\times O(20)$, and as a result they become, after gauge fixing, twisted under the 
$U(1)^2\subset O(4)$ but not under the $U(1)^{10}\subset O(20)$.
The charges of the gravitini in the $(\mathbf{2},\mathbf{1},\mathbf{1})$ representation under $U(1)^2\subset O(4)$
are $(q_1,q_2)=(1/2,1/2)$ while those of the gravitini in the  $(\mathbf{1},\mathbf{2},\mathbf{1})$
representation under $U(1)^2\subset O(4)$
are $(q_1,q_2)=(1/2,-1/2)$. This then results in the mass formul\ae{}~(\ref{fermmas1},\ref{fermmas2}) on using~(\ref{masses}). 

Similarly, the masses of the spin-$1/2$ fields can be found by calculating  the tensors appearing in the mass 
formul\ae{}~(\ref{fermmass}), or by finding the twists in the gauge-fixed theory. Here we do the latter.
Under $SU(2)
\times SU(2)\times O(20)$, the spin-$1/2$ fields
transform as
$$
3 \times (\mathbf{2},\mathbf{1},\mathbf{1})+ 3 \times(\mathbf{1},\mathbf{2},\mathbf{1})  +(\mathbf{2},\mathbf{1},\mathbf{20})+ (\mathbf{1},\mathbf{2},\mathbf{20}) \, . $$
The fermions in the $3 \times (\mathbf{2},\mathbf{1},\mathbf{1})$ representation will all get mass $m_1$ 
given by~(\ref{fermmas1}) while those in the $3 \times(\mathbf{1},\mathbf{2},\mathbf{1}) $
representation will all get mass $m_2$ given by~(\ref{fermmas2}). The remaining fermions will all be massive for generic angles.

We see that something special happens if $\theta_1 =  \theta _2$ and $\tilde\theta_1 = \tilde\theta _2$
so that $m_2$ is zero or $\theta_1 = - \theta _2$ and $\tilde\theta_1 =- \tilde\theta _2$
so that $m_1$ is zero.
In either case, there are 2 massless gravitini and six massless spin-half fields. In this case the vacuum 
breaks the $\mathcal{N}=4$ local supersymmetry to
$\mathcal{N}=2 $ supersymmetry, and the massless fields fit into the
 $\mathcal{N}=2$ supergravity multiplet with three massless vector multiplets, which is  just the spectrum of the STU model. Unlike 
most occurrences of the STU model in string theory, in the present case it does not occur as a truncation of a 
richer  theory, but describes the whole low-energy sector of the theory. 

For generic angles, both $m_1$ and $m_2$ are non-zero therefore all the fermions become massive and all supersymmetry is broken.

For the ungauged $\mathcal{N}=4$  theory, there is a local $H_s$ symmetry and a global $O(6,22)$ symmetry. A monodromy in $O(4)\times O(20) \subset O(6,22)$  will break the global $O(6,22)$ to $O(2,2)$.
However, $O(4)$ has a subgroup $SO(3)_1\times SO(3)_2$, and if the $O(4)$  monodromy is restricted to be in $SO(3)_2$, then $SO(3)_1$ survives as a symmetry in the gauged supergravity. This corresponds to the case $m_1=0$ above.
Similarly, $m_2=0$  corresponds to a monodromy in $SO(3)_1$ with $SO(3)_2$ surviving as a symmetry.
In the formalism with  the local $H_s$ fixed,  there is an $SU(4)\times O(22)$ global symmetry,
of which $SU(4)$ is an R-symmetry. The  monodromies are in $SU(2)_1 \times SU(2)_2 \times O(20)\subset SU(4) \times O(20)$. 
If $m_1=0$, then the monodromies lie 
in $SU(2)_2 \times SO(20)  \subset SU(2)_1 \times SU(2)_2 \times SO(20)$  and hence 
$SU(2)_1$ survives as an R-symmetry in the low-energy theory, and similarly if $m_2=0$ then 
the monodromies lie in $SU(2)_1\times SO(20)  \subset SU(2)_1 \times SU(2)_2 \times SO(20)$  and hence 
$SU(2)_2$ survives as an R-symmetry in the low-energy theory. The surviving $SU(2)$  is the R-symmetry
for the unbroken $\mathcal{N}=2$ supersymmetry.

\subsection{Aspects of the low-energy $\mathcal{N}=2$ theory}
\label{subsec:low}

In this subsection we give further details of  how the $\mathcal{N}=4$ multiplets of the $D=4$ supergravity decompose into 
massless and massive multiplets of $\mathcal{N}=2$ supersymmetry for the cases in which $m_1$ is zero or $m_2$ is zero, 
and study aspects of the effective $\mathcal{N}=2$ theory valid 
at energies much less than the supersymmetry breaking scale set by the gravitini masses.

\subsubsection{Massless multiplets}

The fields in the $D=4$  supergravity theory fit into an $\mathcal{N}=4$ supergravity supermultiplet and $22$
$\mathcal{N}=4$ vector supermultiplets. 
Once the local $SU(4)\times O(22)$ symmetry has been fixed, all fields transform under rigid
$SU(4)\times SO(22)$  transformations. We now give the representations of 
the component fields under this global $SU(4)\times SO(22)$, and the decomposition of these representations into
$SU(2)_1 \times SU(2)_2 \times SO(20) \subset SU(4)\times SO(2))$ representations, which will be useful for studying 
the $\mathcal{N}=2$ multiplet structure. The $\mathcal{N}=4$ supergravity multiplet is
\begin{equation}
\begin{array}{|c|c|c|}
\hline
& SU(4)\times SO(22)& SU(2)_1 \times SU(2)_2 \times SO(20) \\
\hline
2 & (\mathbf{1},\mathbf{1})&(\mathbf{1},\mathbf{1},\mathbf{1})\\
3/2 & (\mathbf{4},\mathbf{1})&(\mathbf{2},\mathbf{1},\mathbf{1})+(\mathbf{1},\mathbf{2},\mathbf{1}) \\
1 &  (\mathbf{6},\mathbf{1})&2\times (\mathbf{1},\mathbf{1},\mathbf{1})+(\mathbf{2},\mathbf{2},\mathbf{1})\\
1/2 & (\mathbf{4}',\mathbf{1})& (\mathbf{2},\mathbf{1},\mathbf{1})+  (\mathbf{1},\mathbf{2},\mathbf{1})\\
0 & 2 \times (\mathbf{1},\mathbf{1})& 2 \times (\mathbf{1},\mathbf{1},\mathbf{1})\\
\hline
\end{array}
\end{equation}
while for the 22 vector multiplets we have

\begin{equation}
\begin{array}{|c|c|c|}
\hline
& SU(4)\times SO(22)&SU(2)_1 \times SU(2)_2 \times SO(20) \\
\hline
1 &  (\mathbf{1},\mathbf{22})&(\mathbf{1},\mathbf{1},\mathbf{20})+2 \times (\mathbf{1},\mathbf{1},\mathbf{1})\\
1/2 & (\mathbf{4},\mathbf{22})&(\mathbf{2},\mathbf{1},\mathbf{20})+ (\mathbf{1},\mathbf{2},\mathbf{20})
+2 \times (\mathbf{2},\mathbf{1},\mathbf{1})+2\times (\mathbf{1},\mathbf{2},\mathbf{1})\\
0 &  (\mathbf{6},\mathbf{22})&  (\mathbf{2},\mathbf{2},\mathbf{20}) + 2 \times (\mathbf{1},\mathbf{1},\mathbf{20}) + 
2\times (\mathbf{2},\mathbf{2},\mathbf{1})+ 4\times (\mathbf{1},\mathbf{1},\mathbf{1}) \\
\hline
\end{array}
\end{equation}
For the gravitini and spin-1/2 fields, the representation $\mathbf{4}$ corresponds to left-handed fermions 
transforming in the $\mathbf{4}$ and right handed ones transforming in the $\bar{\mathbf{4}}$, i.e. 
$\mathbf{4} \sim \bar{\mathbf{4}}_R + \mathbf{4}_L$. Similarly, $\mathbf{4}' \sim \bar{\mathbf{4}}_L + \mathbf{4}_R$. 

If $m_1=0$, then the monodromies lie 
in $SU(2)_2 \times SO(20)  \subset SU(2)_1 \times SU(2)_2 \times SO(20)$ and 
$SU(2)_1$ survives as an R-symmetry in the low-energy theory.  The massless states are the ones that are
singlets under $SU(2)_2 \times SO(20)$, i.e.
\begin{equation}
\begin{array}{|c|c|}
\hline
2     &    (\mathbf{1},\mathbf{1},\mathbf{1})\\
3/2  &      (\mathbf{2},\mathbf{1},\mathbf{1})\\
1 &        4 \times (\mathbf{1},\mathbf{1},\mathbf{1}) \\
1/2 &     3\times (\mathbf{2},\mathbf{1},\mathbf{1})\\
0 &  6\times (\mathbf{1},\mathbf{1},\mathbf{1}) \\
\hline
\end{array}
\end{equation}
This gives $\mathcal{N}=2$ supergravity with three massless $\mathcal{N}=2$ vector multiplets, which is precisely the content of 
the STU model.

\subsubsection{Massive multiplets}

In generic models, the remaining states now organize themselves in massive multiplets. 
The massive states that are singlets under SO(20) are:
\begin{equation}
\begin{array}{|c|c|}
\hline
3/2  &      (\mathbf{1},\mathbf{2},\mathbf{1})\\
1 &         (\mathbf{2},\mathbf{2},\mathbf{1}) \\
1/2 &     3\times (\mathbf{1},\mathbf{2},\mathbf{1})\\
0 &  2\times (\mathbf{2},\mathbf{2},\mathbf{1}) \\
\hline
\end{array}
\end{equation}
This gives a BPS gravitino multiplet and two BPS hypermultiplets. 

The remaining fields from the original $\mathcal{N}=4$ ungauged theory are all in a $\mathbf{20}$ of $SO(20)$, namely:
\begin{equation}
\begin{array}{|c|c|}
\hline
1 &         (\mathbf{1},\mathbf{1},\mathbf{20}) \\
1/2 &     (\mathbf{2},\mathbf{1},\mathbf{20})+(\mathbf{1},\mathbf{2},\mathbf{20})\\
0 &  (\mathbf{2},\mathbf{2},\mathbf{20})+ 2 \times (\mathbf{1},\mathbf{1},\mathbf{20}) \\
\hline
\end{array}
\end{equation}   
States in the $(\mathbf{2},\mathbf{20})$ of $SU(2)_2 \times SO(20)$ 
form a BPS hypermultiplet  and states in the $(\mathbf{1},\mathbf{20})$  give a 
BPS massive vector multiplet (one scalar gets eaten by the vector). 

\subsubsection{Accidental massless multiplets}

In certain models, a fraction of the BPS hypermultiplets are neutral under the monodromy and are therefore massless. 
From the mass formula~(\ref{masses}),  a supergravity field with charges $q_i
$ $(i,j=1,\dots 12$)  under $U(1)^{12}$
will be massless if 
 \begin{equation}
 \label{vanmas1}
 \sum _{i=1}^{12}q_i \theta_i =0    
  \end{equation}
 and 
 \begin{equation}
  \label{vanmas2}
\sum _{i=1}^{12}q_i  \tilde \theta_i =0 \, .
 \end{equation}
For example, for the scalars with mass (\ref{massess})
this will be the case if
$ \theta_M =-\theta_A$ and $\tilde  \theta_M=- \tilde \theta_A$.

\subsubsection{Further accidental massless multiplets from KK modes}

There can be further accidental massless multiplets from Kaluza-Klein  
modes~\cite{Dabholkar:2002sy}. For a trivial reduction without monodromy on the two circles 
with coordinates $y_1,y_2$, each field has a mode expansion of the form
\begin{equation}
\phi (x^\mu, y^1, y^2)=\sum _{n_1,n_2} e ^{i n_1y^1/R_1 +i n_2y^2/R_2}\phi_{n_1,n_2} (x)
 \end{equation}
 with a sum over integers $n_1, n_2$.
 The mode $ \phi_{n_1,n_2} (x)$ then has mass $m$ with
 \begin{equation}
\label{KKmasses}
m^2 = 
\left( 
\frac {n_1 }{ R_1}
\right) ^2 + \left( 
\frac {n_2 }{    R_2}
\right)^2 \, .
\end{equation}

For a reduction with duality twists of the type discussed above, this formula is modified 
 for fields that are charged under $U(1)^{12}\subset O(4)\times O(20)$.
For a mode $ \phi_{n_1,n_2} (x)$  with charges $q_i
$ $(i,j=1,\dots 12$)  under $U(1)^{12}$, 
the mass $m$ will be given by the following modification of~(\ref{masses}):
\begin{equation}
\label{massest}
m^2 = 
\left(
\frac {2\pi n_1 +\sum _{i=1}^{12}q_i \theta_i }{ 2\pi R_1}
\right) ^2 + 
\left(
\frac {2\pi n_2+ \sum _{i=1}^{12}q_i  \tilde \theta_i }{  2\pi R_2}
\right)^2\, .
\end{equation}
In the truncated supergravity theory, the condition for massless states 
were~(\ref{vanmas1}) and~(\ref{vanmas2}).
Now we see that the condition that the full Kaluza-Klein spectrum contains massless modes is that
\begin{equation}
  \label{vanmas3}
 \sum _{i=1}^{12}q_i \theta_i =0     \mod  2\pi
  \end{equation}
 and 
\begin{equation}
 \label{vanmas4}
\sum _{i=1}^{12}q_i  \tilde \theta_i =0   \mod  2\pi \, .
 \end{equation}

For  the hypermultiplets $\phi _{MA}$, the condition that there be a massless KK mode is that 
\begin{equation}
 \theta_M+ \theta_A=0    \mod   2\pi
  \end{equation}
 and 
\begin{equation}
\tilde  \theta_M+ \tilde \theta_A =0  \mod   2\pi \, ,
 \end{equation}
while for 
$\rho_{MA}$ the condition is 
\begin{equation}
 \theta_M- \theta_A=0     \mod  2\pi
  \end{equation}
 and 
\begin{equation}
\tilde  \theta_M- \tilde \theta_A =0   \mod   2\pi \, .
 \end{equation}

For the gravitini, there is a similar modification of the mass formul\ae.
The gravitini KK modes will include massless spin-3/2 fields  
if
$$
\theta_1 + \theta _2=0    \mod   4\pi
 $$
 and 
$$\tilde\theta_1 + \tilde\theta _2 =0   \mod   4\pi \, ,
$$
or
$$
\theta_1 - \theta _2=0    \mod   4\pi
 $$
 and 
$$\tilde\theta_1 - \tilde\theta _2 =0  \mod   4\pi \, .
$$
These are for gravitini modes that are periodic in $y^1, y^2$; the conditions for anti-periodic ones would be slightly different.


\section{Compactifications with non-geometric monodromies}

We will now apply the supergravity framework developed in the last section to the non-geometric compactifications analysed 
in sections~\ref{sec:math} and~\ref{sec:nongeom} from the algebraic geometry and string theory viewpoints. 
In string theory, the non-compact symmetry groups $O(4,20)$ and $O(6,22)$ are broken to the discrete subgroups preserving the 
charge  lattice~\cite{Dabholkar:2002sy}. In particular, $O(4,20)$ is broken to the  group $O(\Gamma_{4,20})$ preserving the lattice 
$\Gamma_{4,20}$, and we choose the natural  basis in which the metric $\eta_{IJ}$
is the metric on the lattice $\Gamma_{4,20}$ given in section~\ref{sec:math}.
As a result, the mass parameters introduced in the twisted reduction now
take discrete values.
Our aim here is to find the 
non-geometric type IIA compactifications, consisting of K3 fibrations over two-tori with non-geometric twists, 
in the sense of~\cite{Dabholkar:2002sy}, that at fixed points of the twist reproduce the orbifold constructions of~\cite{Israel:2013wwa} 
summarized in section~\ref{sec:nongeom}. 

We start with a $(p_1,p_2)-${\it cyclic} K3 surface, $i.e.$ a hypersurface in 
a weighted projective space defined by a polynomial of the form 
\begin{equation}
W = x_1^{\, p_1} + x_2^{\, p_2} + f(x_3,x_4)\, ,
\end{equation}
where $p_1$ and $p_2$ are prime numbers. As we have seen, such surface admits two non-symplectic 
automorphisms $\sigma_{p_1}$ and $\sigma_{p_2}$ 
generating an automorphism group isomorphic to $\mathbb{Z}_{p_1} \times \mathbb{Z}_{p_2}$. 
Using Definition~\ref{def:doub} one can associate to them non-geometric automorphisms  $\hat{\sigma}_{p_1},\hat{\sigma}_{p_2}$
generating a subgroup $O(\hat{\sigma}_{p_1}) \times O(\hat{\sigma}_{p_2})$ of the duality group $O(\Gamma_{4,20})$, 
isomorphic to $\mathbb{Z}_{p_1} \times \mathbb{Z}_{p_2}$; see eqs.~(\ref{sighat},\ref{eq:osigmap}).

In subsection~\ref{subsec:free} we   defined orbifold compactifications consisting of identifying the IIA superstring theory 
on $K3\times T^2$ under the action of $\hat{\sigma}_{p_1}$ combined with a shift of $2\pi R_1/p_1$ for the first one-cycle 
of the torus, and $\hat{\sigma}_{p_2}$ combined with a shift of $2\pi R_2/p_2$ for the second one-cycle of the torus. 
This is all defined at a particular point in the moduli space of CFTs on $K3$ that is a fixed point under these transformations, 
corresponding to the $(p_1,p_2)$-{\it cyclic} K3 surface at a Landau-Ginzburg point.

The reduction with a duality twist  construction gives  a way to extend this to all points in moduli space, 
and then  the supergravity analysis of section~\ref{sec:gauge} gives the  resulting low energy effective field theory.
The twisted reduction gives a fibration of K3 surfaces over $T^2$   with two non-geometric monodromies 
in $O(\Gamma_{4,20})$ associated with the one-cycles of the torus:
\begin{itemize}
\item An order $p_1$ monodromy belonging to $O (\hat{\sigma}_{p_1})$, associated with the non-geometric 
automorphism $\hat{\sigma}_{p_1}$, for the first one-cycle of the torus.
\item An order $p_2$ monodromy belonging to $O (\hat{\sigma}_{p_2})$, associated with 
the non-geometric automorphism $\hat{\sigma}_{p_2}$,  for the second one-cycle of the torus. 
\end{itemize}
The action of $\hat{\sigma}_p$ is deduced from the action of the geometrical automorphism $\sigma_{p}$ on the 
vector space generated by $T(\sigma_{p})$ and from the action of the automorphism $\sigma_{p}^T$ of the mirror surface on the 
vector space generated by $T(\sigma_{p}^T)$, see eqs.~(\ref{sighat}). 
These give the monodromies and hence the structure constants of the associated gauged supergravity.

\begin{exa}\normalfont 
In example~\ref{exa:selfmirror} we constructed an explicit example of an order three non-geometric automorphism that leaves invariant the self-mirror K3 surface~(\ref{eq:self_mirr}) at the Gepner point. 
From ~(\ref{Mblock}), the corresponding $O(\Gamma_{4,20}) $ element is given by the  $24\times 24$ matrix $\hat{M}_3 $
 \begin{equation}
\label{Mblocka}
\hat{M}_3 := \left( \begin{array}{cccc} M_3^{E_8} & 0 & 0 & 0 \\0 & M_3^{U\oplus U} & 0 & 0 \\ 0& 0  & M_3^{E_8}& 0\\
0&0&0&M_3^{U \oplus U} \end{array} \right)\, ,
\end{equation}
where $M_3^{E_8}$ is given by eq.~(\ref{eq:j3e8}) and $M_3^{U\oplus U}$ by eq.~(\ref{eq:j3uu}).
The matrix $M_3^{U\oplus U}$ has   eigenvalues $\exp \frac{2i\pi}{3}$ and   $\exp \frac{4i\pi}{3}$ with degeneracy two for each, 
and the matrix $M_3^{E_8}$ has the eigenvalues $\exp \frac{2i\pi}{3}$ and $\exp \frac{4i\pi}{3}$ each 
with degeneracy four.

The corresponding twisted reduction on $K3\times T^2$ is obtained using the formalism presented 
in subsection~\ref{subsec:twisted_red}. Suppose we reduce on the $y_1$ circle with monodromy
 \begin{equation}
e^{2\pi R_1 N_1}= \hat{M}_3
\end{equation}
This can be put in the form~(\ref{monods}) by a change of basis, with the twelve
 angles $\theta_i $ given by  $2\pi/3$ with degeneracy 16 and $4\pi/3$ with degeneracy 8. 
From this, one can find $N_1$ which in this basis takes the form~(\ref{angless}). 
Then from~(\ref{eq:structconst_def}) the structure constants $t_{iI}{}^J$ are given by
\begin{equation}t_{iI}{}^J=N_{iI}{}^J \, . \end{equation}

Similarly, the order seven non-geometric automorphism would give a 
monodromy matrix $\hat{M}_7$ which can be brought to the diagonal form~(\ref{monods}) with the twelve
 angles $\theta_i $ given by $\exp (2 i r
\pi/7)$, for $r=1,...,6$, each with degeneracy 4.

To specify the reduction,
we choose the monodromy $e^{2\pi N_2}$ for the other circle from another automorphism, $e.g.$ 
that resulting from $\sigma_3 $ or $\sigma_7$, and this gives the structure constants  $t_{2I}{}^J$.
\end{exa}

In full generality, using Lemma~\ref{lem:diag} in section~\ref{sec:math}, the $GL(24;\mathbb{Z})$ matrices associated with the non-geometric automorphisms can be diagonalized over $\mathbb{C}$, or equivalently 
can be written as elements of $O(4,20)$, the group preserving the Minkowski metric $\text{diag}\, (1_4,-1_{20})$, by a change of basis. The monodromies in $\mathbb{Z}_{p_1} \subset O(4)\times O(20)\subset O(4,20)$ can be brought to 
the standard form (\ref{monods}) specified by 12 angles $\theta _i$  which satisfy $\exp (i \theta _i p_1)=1$.\footnote{More 
explicitly, there exists a positive integer $q$ such that the complex numbers $\{ \exp i \theta_i, i=1,\ldots,12 \}$ are given by the 
primitive $p$ roots of unity, $q$ times each.} In the same way the monodromies in $\mathbb{Z}_{p_2} \subset O(4)\times O(20)\subset O(4,20)$ are specified by 12 angles $\tilde \theta _i$  which satisfy $\exp (i\tilde \theta _i p_2) =1$.

From these angles, the full structure of the effective supergravity theory can be read off, as seen in 
section~\ref{sec:gauge}. The scalar potential of the supergravity admits  minima at the fixed points of the 
automorphisms that reproduce the four-dimensional physics obtained from the asymmetric Gepner models considered 
in section~\ref{sec:nongeom}.  

The stringy compactifications discussed in this work have Minkowski vacua that preserve $\mathcal{N}=2$ 
supersymmetry as we will show below, and have three massless vector multiplets S,T and U  
associated respectively to the axion-dilaton and to the $T^2$ moduli. 
About half of the corresponding asymmetric Gepner models, for instance the self-mirror surface~(\ref{eq:self_mirr}) with $\hat{\sigma}_3$ 
and $\hat{\sigma}_7$ monodromies, give just $\mathcal{N}=2$ STU supergravity at low energies, while in the other cases 
the low-energy theory contains some additional massless hypermultiplets, depending of the choice of K3 surfaces and of 
automorphisms; the associated moduli space will be discussed in subsection~\ref{subsec:mod_space}.

\subsection{Gravitini masses and supersymmetry}

As we have seen, the  isometry induced by the action of the non-geometrical automorphism $\hat{\sigma}_p$ 
generates a finite order subgroup conjugate to a subgroup of 
$[O(2)\times O(20-\rho_p)] \times [O(2)\times O(\rho_p)]$. 
As far as gravitini masses are concerned, only the space-like subgroup 
$O(2) \times O(2) \subset O(4) \subset O(4,20)$ 
plays a role, where the first $O(2)$ factor acts as an order $p$ rotation in the space-like two-plane in  the vector space 
generated by $T (\sigma_p)$ while the second $O(2)$ factor acts as an order $p$ rotation in the space-like 
two-plane in  the vector space generated by $T (\sigma_p^T)$.\footnote{For instance, for the order three automorphism 
studied in example~\ref{exa:selfmirror}, each  $O(2)$ generator comes from the $O(2,2;\mathbb{Z})$ generator 
given by eq.~(\ref{eq:j3uu}) after $O(2,2;\mathbb{R})$ conjugation.}

The parts of the monodromies in  $O(2)\times O(2) \subset O(4)$ transformations are specified by the angles
$\theta_1,\theta_2$ and $\tilde \theta_1, \tilde \theta_2$.
These are then
\begin{equation}
 \theta_1 = \frac {2\pi} {p_1}
 , \qquad
 \theta_2=\varepsilon_1  \frac {2\pi} {p_1}
\end{equation}
and
\begin{equation}
\tilde \theta_1 = \frac {2\pi} {p_2}
 , \qquad
\tilde \theta_2=\varepsilon_2  \frac {2\pi} {p_2} \, .
\end{equation}
Here $\varepsilon_i=0$ if there is no discrete torsion.
If there is discrete torsion, then 
$\varepsilon_i \in\{-1 , 1\}$ corresponding to the two possible choices  of discrete torsion for each cycle,
as seen from  the corresponding worldsheet description in subsection~\ref{subsec:autLG}.

As explained in section~\ref{sec:gauge}, $\mathcal{N}=2$ supersymmetry is preserved only 
 if $\theta_1 =  \theta _2$ and $\tilde\theta_1 = \tilde\theta _2$
 or $\theta_1 = - \theta _2$ and $\tilde\theta_1 =- \tilde\theta _2$. This requires $\varepsilon_1 = \varepsilon_2 =  1$ or $\varepsilon_1 = \varepsilon_2 = - 1.$ 
Otherwise all supersymmetry is broken. Note that accidental supersymmetry from KK modes cannot arise here if $p_1>2$ or $p_2>2$.

We then draw the following conclusions which are in accord with the Gepner model description of the vacua that 
was obtained in~\cite{Israel:2013wwa} (see in particular around eq.~(4.3) in~\cite{Israel:2013wwa}):
\begin{itemize}
\item For \lq geometric' non-symplectic automorphisms of K3 surfaces,  
which have vanishing discrete torsion $\varepsilon_i=0$,  
 all gravitini become massive and so all the spacetime supersymmetry is broken. 
\item The two non-geometric twists preserve the same $\mathcal{N}=2\subset \mathcal{N}=4$ space-time supersymmetry if $\varepsilon_1 = \varepsilon_2 = \pm 1$. Then 
two of the gravitini remain massless, while the other two acquire an equal mass. In the worldsheet description, 
$\varepsilon_1 = \varepsilon_2$ means that 
the same choice of discrete torsion was made for both of the corresponding Gepner model freely acting orbifolds, 
so that they both preserve space-time supercharges from either the left-movers or the right-movers. 
\item Whenever $\varepsilon_1 = \varepsilon_2 = \pm 1$ the isometry preserves $SO(3) \subset SO(4)$ and  hence an $SU(2)_R \subset SU(4)_R$ of the space-time R-symmetry is preserved.  
\end{itemize}
Then only the non-geometric twists with discrete torsion that pair the non-symplectic  
automorphisms with the corresponding automorphisms acting on the mirror K3 surfaces can be compatible with 
$\mathcal{N}=2$ vacua in four dimensions. 

To conclude,  there is a perfect agreement between the   gauged $\mathcal{N}=4$ supergravity and the worldsheet construction.  Note finally 
that the mass scale of the spontaneous supersymmetry breaking $\mathcal{N}=4 \to \mathcal{N}=2$ is set by the (inverse of the) 
volume of the two-torus~\cite{Israel:2013wwa}  and   can be taken to be much smaller than the string mass scale. 
Therefore it makes sense to analyse the model  within the four-dimensional supergravity framework (while  ten-dimensional supergravity 
would be inappropriate for these non-geometric constructions).


\subsection{Moduli space}
\label{subsec:mod_space}

The mathematical formulation of the non-geometric automorphisms that we have given in 
section~\ref{sec:nongeom} provides precise predictions for the scalar manifolds arising from the moduli space 
of our models, parametrized by the vacuum expectation values of the accidental massless hypermultiplets 
discussed in subsection~\ref{subsec:low} from a supergravity viewpoint.

From the general construction of compactifications with duality 
twists~\cite{Dabholkar:2002sy},  the minima of the effective four-dimensional 
potential~(\ref{potential}) correspond to the intersection of the fixed-point loci of the 
two monodromies used in the reduction. Hence the remaining 
massless hypermultiplets, if any, correspond to K3 moduli that are invariant under both 
automorphisms $\hat{\sigma}_p$ used in a particular compactification.

For a given non-geometric automorphism $\hat{\sigma}_p$, these moduli arise first as deformations of the algebraic 
surface that preserve its $p$-cyclic form~(\ref{eq:pcycl}), $i.e.$ such that the surface admits the action of the 
non-symplectic automorphism $\sigma_p$. 
The global structure of these moduli spaces was studied in~\cite{Dolgachev2007} and can be summarized briefly as follows 
(for the details see \cite{Dolgachev2007}). 
The complex vector space $T(\sigma_p)^{\mathbb{C}}:= T(\sigma_p)\otimes \mathbb{C}$, generated by the orthogonal 
complement of the invariant lattice $S(\sigma_p)$, admits a decomposition in terms of the eigenspaces of 
$\sigma_p^\star$, see lemma~\ref{lem:diag}. One has from~(\ref{diago})
\begin{equation}
T(\sigma_p)^{\mathbb{C}} = T_{\zeta_p}\oplus T_{\zeta_p^{\, 2}}\cdots 
\oplus  T_{\zeta_p^{\, p-1}}\, .
\end{equation}
Following~\cite{Dolgachev2007}, let us define 
\begin{equation}
\mathcal{B}_p= \{ z \in\mathbb{P} (T_{\zeta_p} ) \, , \ (z,z) = 0, (z,\bar z) >0 \} \, , 
\end{equation}
where $\mathbb{P}(T_{\zeta_p} )$ denotes the projective space associated with the complex vector space $T_{\zeta_p}$ and $(-,-)$ denotes the bilinear form on the K3 lattice $\Gamma_{3,19}$ that induces in a natural way a bilinear form on $T(\sigma_p)$.\footnote{Recall that given a K3 surface $X$,  a {\it marking} is the choice of an isometry $\phi: H^2(X,\IZ)\lra \Gamma_{3,19}$  and this extends in a natural way to $\phi_{\IC}: H^2(X,\IC)\lra \Gamma_{3,19}\otimes \IC$.
Then if $\omega(X)$ is the holomorphic 2-form we have $H^{2,0}(X)=\IC\omega(X)$ and $\phi_{\IC}(H^{2,0}(X)$ is a point of
$$
\Omega_{K3}=\{[\sigma]\in\IP(\Gamma_{3,19}\otimes \IC)|(\sigma,\sigma)=0, (\sigma,\bar\sigma)>0\}
$$
which is an open (analytic) subset in a 20-dimensional quadric of the 21-dimensional projective space $\IP(\Gamma_{3,19}\otimes \IC)$. The point $\phi_{\IC}(H^{2,0}(X)$ is called {\it period point of the marked K3 surface} and the moduli space of marked K3 surfaces is a quotient of $\Omega_{K3}$.}
We define also 
\begin{equation}
\Gamma_p = \{ \gamma \in O(T(\sigma_p))\, ,\  \gamma \circ \sigma_p^\star = \sigma_p^\star \circ \gamma \}\, , 
\end{equation}
the subgroup of the isometry group $O(T(\sigma_p))$ commuting with the action of $\sigma_p$.

Then the K3 surface with non-symplectic automorphism $\sigma_p$ has period (i.e. holomorphic two form $\omega(X)$) 
lying in  the following space:
\begin{equation}
\label{eq:mod_fix}
 \mathcal{M}_{\textsc{cs}\, ,  \textsc{Fix}}^{p} \cong \Gamma_p \backslash \mathcal{B}_p \, .
\end{equation}
For $p>2$, $\mathcal{B}_p$ is of complex dimension $\rk (T(\sigma_p))/(p-1)-1$ and is isomorphic to a complex ball.\footnote{For the self-mirror surface, for instance, 
local coordinates on the moduli space  $\Gamma_p \backslash \mathcal{B}_p$ are given by the 
monomial deformations of~(\ref{eq:self_mirr}) that are invariant under the action of $\sigma_p$. For the automorphism $\sigma_3$ one gets the monomials
$\{ y^n z^{42-6n}, n=1,\ldots,5 \}$, generating a space of complex dimension five.} For $p=2$ one gets a Hermitian symmetric space of complex dimension $\rk (T(\sigma_p))-2$.

We now consider the non-geometric   automorphism $\hat{\sigma}_p$ constructed from $\sigma_p$ as defined in 
section~\ref{sec:nongeom}. To understand its action on the CFT moduli space one has to look also at 
the mirror surface $X_{W^T, G^T}$ which admits an action of the non-symplectic 
automorphism $\sigma_p^T$.  In the same way as before, we define\footnote{Recall that $S^{\mathfrak{Q}}(\sigma_p)\cong T(\sigma_p^T)$.}
\begin{equation}
T(\sigma_p^T)\otimes \mathbb{C} = T_{\zeta_p}^T \oplus T_{\zeta_p^{\, 2}}^T \cdots \oplus  T_{\zeta_p^{\, p-1}}^T\, ,
\end{equation}
\begin{equation}
\mathcal{B}_p^T= \{ z \in\mathbb{P} (T_{\zeta_p}^T ) \, , \ (z,z) = 0, (z,\bar z) >0 \} \, , 
\end{equation}
and 
\begin{equation}
\Gamma_p^T = \{ \gamma \in O(T(\sigma_p^T))\, ,\  \gamma \circ (\sigma_p^T)^\star = (\sigma_p^T)^\star \circ \gamma \}\, .
\end{equation}
The moduli space of K3 surfaces with non-symplectic action by $\sigma_p^T$ is then given by $\Gamma_p^T \backslash \mathcal{B}_p^T$. 

To summarize,  by using the description of the period map for K3 surfaces, the K3 surface with 
non-symplectic automorphism $\sigma_p$ has period in $\Gamma_p \backslash B_p$, and the mirror K3 surface 
has period in $\Gamma_p^T \backslash B_p^T$. By using the definition of $\hat\sigma_p$ (see Definition~\ref{def:doub}),  
we expect that moduli space of CFTs on $S(\sigma_p)$-polarized K3 surfaces that are invariant under the action of the non-geometric automorphism 
$\hat{\sigma}_p$ is obtained by the direct product of these two spaces:
\begin{equation}
\label{eq:mod_fix_nongeom}
\hat{\mathcal{M}}_{\Sigma\, ,  \textsc{Fix}}^{p} \cong  \Gamma_p \backslash \mathcal{B}_p \ \times \ \Gamma_p^T \backslash \mathcal{B}_p^T \, , 
\end{equation}
and is of complex dimension (recall that we are considering here $p\in \{3,5,7,13 \}$):
\begin{equation}
\text{dim}_\mathbb{C}\, \left(\hat{\mathcal{M}}_{\Sigma\, ,  \textsc{Fix}}^{p}\right) = \left\{ 
\begin{array}{lr}
\frac{24}{p-1} -2 &\ , \quad p>2 \, ,\\
20 &\ , \quad p=2\, . 
\end{array}\right.
\end{equation} 
Interestingly, this dimension is the same for every automorphism $\sigma_p$ of a given prime order $p$, regardless of the rank of the 
corresponding invariant lattice $S(\sigma_p)$. The full CFT moduli space, that corresponds to the hypermultiplet moduli space 
in supergravity, has twice this dimension as was checked from the string theory spectra computed in~\cite{Israel:2013wwa}.

With two monodromy twists associated with the two one-cycles of the two-torus, one should consider the intersection of the corresponding 
moduli spaces, which is easier to do case by case.  For instance, for the 
self-mirror K3 surface~(\ref{eq:self_mirr}) twisted by the non-geometric monodromies $\hat{\sigma}_3$ and  $\hat{\sigma}_7$, this intersection is just a point\footnote{In fact the moduli space of 
K3 surfaces with an action by $\sigma_7$ is one-dimensional, and the K3 surfaces of the family carry an elliptic fibration \cite[Example 6.1, 1)]{AST}. One can check that only one K3 surface 
of the family admits also a non-symplectic automorphism of order three, this is then the K3 surface in example (\ref{eq:self_mirr}). This explains why the intersection is only one point.} 
and so there are no  massless hypermultiplets in the low energy  supergravity and we obtain just the $\mathcal{N}=2$ STU supergravity model (provided that $\varepsilon_1 = \varepsilon_2$ so that 
the two automorphisms preserve the same half of the supersymmetry).


\section{Conclusion}

In this work we have constructed a new class of $\mathcal{N}=2$ four-dimensional non-geometric compactifications of type IIA superstring theories, that consist of K3 fibrations over two-tori with non-geometric monodromies which lead in most cases to pure $\mathcal{N}=2$ 
STU supergravity with no hypermultiplets at low energies. 

The monodromies correspond to non-geometric automorphisms that we have obtained 
by pairing a non-symplectic automorphism of a K3 surface with a non-symplectic automorphism of the mirror surface. We 
have demonstrated that the action of such an automorphism can be lifted to an isometry of the lattice $\Gamma_{4,20}$, $i.e.$ 
an element of the duality group  $O(\Gamma_{4,20})$  of CFTs on K3 surfaces, and hence leads to a well-defined string theory compactification. We have shown that the fixed loci of these automorphisms are given by asymmetric Gepner model orbifolds, 
considered recently in~\cite{Israel:2013wwa}. The new understanding of these non-geometric backgrounds in terms of mirrored automorphisms 
should apply to non-geometric automorphisms of Calabi-Yau three-folds as well (except naturally the lattice-related aspects).

We have analysed the models from the four-dimensional $\mathcal{N}=4$ gauged supergravity perspective valid at low energies. 
The matrices corresponding to the $\Gamma_{4,20}$ isometries that we have constructed provide 
directly the structure constants which parametrise the gauged supergravities obtained from twisted reductions of $K3\times T^2$, and 
we have showed that the minima of the superpotential preserve $\mathcal{N}=2$ supersymmetry 
in four dimensions, as expected from the string theory 
constructions of such vacua. 

We plan to provide more details on the four-dimensional gauged supergravity construction in a companion paper. In particular we will 
analyse the scalar manifold of the low-energy theory  in order to show explicitly that the predictions from algebraic geometry are verified, 
and check that all the consistency conditions of gauged supergravity are met for these particular gaugings.

The duality between the type IIA string theory compactified on K3  and  the heterotic string compactified on 
$T^4$~\cite{Hull:1994ys} gives a heterotic dual to our constructions consisting of a toroidal 
reduction of the heterotic string with monodromy twists, that gives an asymmetric orbifold 
construction at fixed points; such models were introduced in~\cite{Dabholkar:2002sy}.
Particular examples of heterotic asymmetric orbifolds are given by CHL 
compactifications~\cite{Chaudhuri:1995fk}; the latter correspond, on the type IIA side, to symplectic automorphisms of $K3$ surfaces. 
Here, algebraic geometry leads us to a particularly interesting class of constructions that 
correspond to non-symplectic $K3$ automorphisms on the type IIA side and preserve $\mathcal{N}=2$ supersymmetry. 
The corresponding heterotic orbifolds will be discussed further in a forthcoming publication.

\subsection*{Acknowledgments}

We thank Samuel Boissi\`ere, Miranda Cheng, Alessandro Chiodo,  Nick Halmagyi, Jan Louis,  Ronen Plesser, Nathan Priddis,  Matthieu Sarkis and Jean-Bernard Zuber  for discussions and correspondence. D.I. research received support from the ILP LABEX (ANR-10-LABX-63) 
supported by French state funds managed by the ANR (ANR-11-IDEX-0004-02) and from the project QHNS in the program ANR Blanc 
SIMI5 of Agence National de la Recherche. 
The  work of CH was supported  by  the EPSRC programme grant 
``New Geometric Structures from String Theory" EP/K034456/1 and the STFC  grant ST/L00044X/1.

\section{Appendix: Explicit lattice computations}
To get the action of the isometry on the lattice $E_8$, respectively $E_6$, with the standard bilinear form $E_i^{st}$, $i=6,8$ as given in Examples \ref{exa:selfmirror}, \ref{exa:autre} we use 
the fact that thanks to the automorphism $\sigma_3$ of order three these two lattices 
have the structure of a $\IZ[\zeta]$-module, where $\zeta$ is a primitive third root of unity (recall that the ring $\IZ[\zeta]$
is called the {\it ring of Eisenstein integers}). Lattices with this property are very much investigated in number theory, see \cite{BF} for a precise introduction of the basic tools needed in this section and a general introduction on the subject. 

Recall that the multiplication of an  element $a+b\zeta \in \IZ[\zeta]$ with an element $x$ in the lattice, is defined as
$$
(a+\zeta b)\cdot x:=ax+b\sigma_3^{\, \star}(x).
$$
Since $\sigma_3^{\, \star}$ by construction does not fix any vector on $E_8$, resp. $E_6$, and $\IZ[\zeta]$ is a principal ideal domain 
then these two lattices are free over $\IZ[\zeta]$ of rank 4, respectively 3. 

Let $e_1,e_2,e_3,e_4$ be a set of generators of the $\IZ[\zeta]$-module $E_8$
so that $E_8=\IZ[\zeta]e_1\oplus\IZ[\zeta]e_2\oplus\IZ[\zeta]e_3\oplus\IZ[\zeta]e_4$ and clearly 
$\mathbb{B}_8=\{e_1,e_2,e_3,e_4, \zeta e_1,\zeta e_2,\zeta e_3,\zeta e_4\}$ is a set of generators of $E_8$ as an integral lattice. 
One can consider a similar set of generators for $E_6$ as $\IZ[\zeta]$-module and a set of generators $\mathbb{B}_6$ for $E_6$ as a lattice over the integers. Following \cite[Chapter 1]{ACT}, consider the hermitian forms on $E_8$, respectively $E_6$ (as $\IZ[\zeta]$-lattices)  
\begin{eqnarray*}
h_{E_8}=\left(\begin{array}{cccc}
3&\theta&0&0\\
\bar\theta&3&\theta&0\\
0&\bar\theta&3&\theta\\
0&0&\bar\theta&3
\end{array}\right),\qquad h_{E_6}=\left(\begin{array}{ccc}
3&\theta&0\\
\bar\theta&3&\theta\\
0&\bar\theta&3\\
\end{array}\right)
\end{eqnarray*}
where $\theta=\zeta-\bar\zeta$.
One can then define a bilinear form  
$$
b_{E_i}(\alpha,\beta):=-\frac{1}{3}(h_{E_i}(\alpha,\beta)+\rho (h_{E_i}(\alpha,\beta))), \, (\alpha,\beta)\in E_i\times E_i
$$
on the lattices $E_i$ with the set of generators $\mathbb{B}_i$, $i=6,8$, where $\rho$ denotes the $\IQ$-automorphism of $\IQ(\zeta)$ that sends $\zeta$ to $\bar\zeta$ (see~\cite{Conway:1987:SLG:39091}). Observe that the element $(\alpha,\beta)$ is considered on the left hand side as an element of the rank $i$ integral lattice and on the right hand side as an element of the rank $i/2$ module over $\IZ[\zeta]$.

With the help of MAGMA one can find a base change matrix $T_{E_i}$ with integer coefficients such that
$$
E_i^{st}=T_{E_i}^t b_{E_i} T_{E_i}, \qquad i=6,8.
$$
where recall that $E_i^{st}$ is the standard lattice metric   as given in the Examples \ref{exa:selfmirror}, \ref{exa:autre} and by abuse of notation
we identify the bilinear form $b_{E_i}$ with its symmetric $i\times i$ matrix. 
The action of the isometry in the above given set of generators $\mathbb{B}_i$ of $E_i$, $i=6,8$ (as an integral lattice) is easy to write,
since this is a block matrix with 4, respectively 3, blocks of the form
\begin{eqnarray*}
\left(\begin{array}{cc}
0&-1\\
1&-1
\end{array}\right).
\end{eqnarray*}
We call these two matrices $H_{E_i}$, $i=6,8$. Then the action
of the isometry in the set of generators with the standard lattice metric is given by
$$
J_3^{E_i}=T_{E_i}^{-1}H_{E_i} T_{E_i}, \qquad i=6,8.
$$
These are the matrices given in the equations \eqref{eq:j3e8}, \eqref{eq:j3e6}.

In a similar way one can compute the action of the automorphism $\sigma_3^{\, \star}$ on $A_2$ and on $U\oplus U$. In this case
the matrices of the hermitian forms have a very easy form
\begin{eqnarray*}
h_{A_2}=(3)\qquad h_{U\oplus U}=\left(\begin{array}{cc}
0&\theta\\
\bar \theta&0\\
\end{array}\right).
\end{eqnarray*}
With the same notation as above we find that in these two cases $T_{A_2}=T_{U\oplus U}=id$, which is not necessarily the case in general
(e.g. one can check in the previous computation that $T_{E_i}\neq id$  since $b_{E_i}\neq E_i^{st}$, for $i=6,8$).

As an example we do the explicit computations for $A_2$. Here recall $$b_{A_2}(\alpha,\beta):=-\frac{1}{3}(h_{A_2}(\alpha,\beta)+\rho (h_{A_2}(\alpha,\beta))), \, (\alpha,\beta)\in A_2\times A_2.$$
As a $\IZ[\zeta]$-lattice we have that $A_2=\IZ[\zeta] e_1$ for a generator $e_1$ so that $\{e_1,\zeta e_1\}$ is a basis of $A_2$ as an integral lattice. Now with $h_{A_2}=(3)$ we have
\begin{eqnarray*}
b_{A_2}(e_1,e_1)=-\frac{1}{3} (h_{A_2}(e_1,e_1)+\rho (h_{A_2}(e_1,e_1)))=-\frac{1}{3}(3+3)=-2,\\
b_{A_2}(\zeta e_1,\zeta e_1)=-\frac{1}{3}(\zeta 3\bar\zeta+\bar\zeta 3\zeta)=-2,\\
b_{A_2} (e_1,\zeta e_1)=b_{A_2} (\zeta e_1,e_1)=-\frac{1}{3}(3\zeta+3\bar\zeta)=1,
\end{eqnarray*}
and we get $b_{A_2}=A_2^{st}$. Now the action of the automorphism $\sigma_3^{\, \star}$ on the basis $e:=e_1$, $f:=\zeta e_1$ of the integral lattice $A_2$ corresponds
by definition of the structure of $\IZ[\zeta]$-module to the multiplication by $\zeta$. So we have that the automorphism sends $e$ to $f$ and 
since $\zeta^2 e_1=-e_1-\zeta e_1$ we get that the image of $f$ is $-e-f$ as given in Example \ref{exa:autre}. In the case of $E_i$, $i=6,8$ we determine with MAGMA a matrix $T_{E_i}$ that changes the basis $\mathbb{B}_i$ to the basis for the standard action and we use then this matrix to get the action of the automorphism on the basis of the standard action. 

Observe that one could use a similar method to determine the action of the automorphism $\sigma_7^{\, \star}$ of Example 
\ref{exa:selfmirror} on the lattice $U\oplus U\oplus E_8$. This is a $\IZ[\zeta_7]$-module of rank 2 ($\zeta_7$ denotes a primitive seventh root of unity) but we do not know the explicit matrix of the hermitian form $h_{U\oplus U\oplus E_8}$ which is a $2\times 2$-hermitian matrix (to determine such forms is in general a difficult problem; see \cite{BF}).  As seen above this would allow us to find the matrix of the base change $T_{U\oplus U\oplus E_8}$ that can then be  used to give the action of $\sigma_7^{\, \star}$ on $U\oplus U\oplus E_8$ with the standard bilinear form.

\bibliography{biblio}

\end{document}